\documentclass[pdftex,twocolumn,epjc3]{svjour3}          

\RequirePackage[T1]{fontenc}

\smartqed  
\usepackage{amsmath,amssymb,amsfonts,multicol}
\RequirePackage{graphicx}
\RequirePackage{mathptmx}      
\RequirePackage{flushend}
\RequirePackage[numbers,sort&compress]{natbib}
\RequirePackage[colorlinks,citecolor=blue,urlcolor=blue,linkcolor=blue]{hyperref}

\begin{document}

\title{Thin accretion disk signatures in  hybrid metric-Palatini gravity}


\author{Polina Dyadina\thanksref{e1,addr1}
        \and
        Nikita Avdeev\thanksref{e2,addr1} 
}

\thankstext{e1}{e-mail: guldur.anwo@gmail.com}
\thankstext{e2}{naavdeev1995@mail.ru}

\institute{Sternberg Astronomical Institute, Lomonosov Moscow State University, Universitetsky Prospekt, 13, Moscow, 119234, Russia\label{addr1}
        }

\date{Received: 13 October 2023 / Accepted: 20 January 2024}

\maketitle

\begin{abstract}
In the present work, accretion onto a static spherically symmetric black hole in the hybrid  metric-Palatini gravity  is considered.  The Novikov-Thorne model for a relativistic thin accretion disk is used. The energy flux, temperature distribution, emission spectrum and energy conversion efficiency of accretion disks around such black holes are numerically calculated.  A comparison with the results for a Schwarzschild black hole is made and conclusions about the viability of the model are drawn. As a result, it is obtained that the accretion disks around black holes in hybrid metric-Palatini gravity are colder and less luminous than in general relativity. 
\end{abstract}

\section{Introduction}\label{sec1}

At the beginning of the XX century,  general relativity (GR) replaced Newton's theory of gravity, offering solutions to long-standing unresolved issues. GR has since remained the generally accepted theory of gravity. However, as physics progressed, new challenges emerged that could not be addressed by GR alone. One of the main problems of theoretical physics at the moment is the accelerated expansion of the universe \cite{Perlmutter1999, Riess1998, Riess2004, Riess2001, Perlmutter1999a}. Another problem manifests itself at the scale of galaxies, demonstrating the presence of a large amount of unknown matter \cite{oort, zwicky}. In addition, there is still no self-consistent theory of quantum gravity. There are problems in the early Universe that cannot be solved within the framework of GR, which leads to the emergence of a large number of inflation theories \cite{Starobinsky1980, Guth1981, Linde1982}. One of the ways to solve these problems is to search for a theory of gravity that go beyond the scope of GR. One of the simplest and most widely used methods for extending GR is f(R)-gravity.

The f(R)-gravity approach allows replacing the Ricci scalar in the Einstein-Hilbert action with an arbitrary  function of the curvature \cite{Bergmann1968, Felice2010, Nojiri2017, Nojiri2011}. An attractive feature of f(R)-gravity is the possibility of simultaneously describing both the inflationary period and the modern accelerated expansion of the universe. The large family of f(R)-theories can be divided into two classes: metric and Palatini ones. In the metric approach the only variable is the metric, while in the Palatini approach an additional variable is the independent affine connection. Within the framework of the metric approach, it is possible to successfully describe the dynamics of the modern universe. However, it encounters difficulties in explaining the dynamics within the solar system \cite{Chiba2003, Olmo2005, Olmo2007}. Nevertheless,  there are a number of viable models that can overcome these difficulties \cite{odintsov1, odintsov2, odintsov3}. Another approach  is the Palatini one, which assumes the affine connection is independent of the metric \cite{Capozziello2007, Sotiriou2010}. However, this modification also faces limitations when it comes to describing observational data \cite{Koivisto2006, koivisto206}.

To overcome these challenges, hybrid metric-Palatini gravity (HMPG) was developed  \cite{Harko2012, Capozziello2015, Harko2020}. This theory combines the advantages of the metric and Palatini approaches while avoids their shortcomings. HMPG includes both metric (Einstein-Hilbert action) and Palatini (arbitrary function of the Palatini curvature) parts. One of the main successes of the model is the  simultaneous explanations for cosmological \cite{Boehmer2013, Lima2016, Capozziello2013} and solar system dynamics \cite{Leanizbarrutia2017, Dyadina2019} without the need for screening mechanisms. Also it provides the description of galactic dynamics \cite{Capozziello2013a}. Moreover, this model has a scalar-tensor representation, simplifying its study. 

HMPG was explored in many works. For the most complete review of studies, see \cite{Harko2020}. In this brief introduction, we outline the main areas of research within the framework of this theory. First of all, it is necessary to note that the HMPG was investigated on a wide range of scales and gravitational regimes. On a cosmological scale, the theory shows its consistency and excellent agreement with observational data \cite{Boehmer2013, Lima2016, Capozziello2013}. On the scale of galaxies, the HMPG allows  to describe rotation curves without introducing a large amount of dark matter \cite{Capozziello2013a}, and to explain the virial mass discrepancy in clusters of galaxies \cite{Capozziello13}. In addition, the theory describes the observational data obtained within the solar system \cite{Leanizbarrutia2017}. It was tested using the parametrized post-Newtonian (PPN) formalism in the weak field regime \cite{Dyadina2019, Dyadina_2022}. HMPG was also studied in   stronger gravitational fields, for example, using observational data from binary  pulsars \cite{Dyadina2018, Avdeev2020}. Gravitational waves were investigated \cite{Dyadina2022, kausar2018}. Moreover, the physical properties of neutron, Bose-Einstein condensate and quark stars were studied \cite{Danil2017}. In the strong field limit,   the static spherically symmetric black hole solution was obtained numerically \cite{Danila2019}. The evolution of dynamic traversable wormhole geometries in a FLRW background in the context of HMPG was studied \cite{Zangeneh2021}. The possibility of obtaining stable spherically symmetric analytical solutions within the framework of this model was also investigated \cite{Bronnikov2020}. Also, in a number of works, a generalized version of the HMPG was considered \cite{Rosa2017, Rosa20, Rosa2021}.

The recently obtained black hole solution \cite{Danila2019} in HMPG has opened up new possibilities for testing this theory. In addition to the fact that a gravitational theory must confirm the existence of compact bodies, it must also predict accretion disks around such objects that are consistent with observations. Therefore, the study of the properties of accretion disks can be used to test the adequacy of both the theory itself and the black hole solution, and can also serve in the future to limit HMPG.

In this paper, our focus is on studying accretion onto a static spherically symmetric black hole within HMPG. Accretion, the process of matter infall onto a black hole, is highly sensitive to the gravitational theory's peculiarities, making it a valuable testbed for exploring new results and constraints for this model. Historically, the first accretion model was created by N. Shakura and R. Sunyaev \cite{Shakura1973}. However, the first model which takes into account  relativistic effects was developed by I.  Novikov, K. Thorne and D. Page \cite{Page1974,Novikov1973}. Previously, many modified theories of gravity have already been investigated by considering accretion disks \cite{Harko2011, mohaddese, Harko2010, Pun2008, Perez2013, Perez2017}.

We employ a numerical solution for the static spherically symmetric black hole metric in HMPG \cite{Danila2019} to calculate the energy flux, luminosity, and temperature within the Novikov-Thorne model. We compare our results with calculations performed for a Schwarzschild black hole in GR. Our analysis considers different values of the background scalar field, first derivative of the scalar field, scalar field mass, and different their interconnections. We examine two cases: one with a Higgs-type potential $V(\phi)=-\frac{\mu^2}{2} \phi^2 + \frac{\zeta}{4}\phi^4$ and another without any potential ($V = 0$). In the conclusion, we discuss the viable aspects of HMPG.

The article is organized into seven sections. The first and last are the introduction and conclusion respectively. The second section provides a description of the HMPG and its scalar-tensor representation. In the third section, we present the Novikov-Thorne model. The fourth section outlines the numerical calculations for the static spherically symmetric black hole metric. In the fifth section the results of numerical calculating for accretion properties in HMPG are presented. In the sixth section we discuss obtained results. The conclusion summarizes our findings.
	
	Throughout this paper the Greek indices $(\mu, \nu,...)$ run over $0, 1, 2, 3$ and the signature is  $(-,+,+,+)$. All calculations are performed in the CGS system.

\section{Hybrid metric-Palatini gravity}\label{sec2}
	In this section, we discuss the fundamental characteristics of HMPG and introduce mathematical background of this theory. The action  is formulated as follows \cite{Harko2012, Capozziello2015}: 
	\begin{equation}
		\label{action1}
		S=\frac{1}{2k^2}\int d^4x\sqrt{-g}[R+f(\Re)]+S_m,
	\end{equation}
where $k^2=\frac{8\pi G}{c^4}$, $G$ is the gravitational constant, $c$ is the speed of light, $g=det\{g_{\mu\nu}\}$~ is the determinant of the metric, $R$ and $\Re$ are the metric and Palatini curvatures respectively and $S_m$~ is the matter action. The first part of the action, the metric curvature $R$, corresponds to the Einstein-Hilbert action. The second part consists of general  function of the Palatini curvature and includes all deviations from GR.  

The general form of scalar curvature expressions is the same:

	\begin{eqnarray}
		\label{R}
		R= g^{\mu\nu}R_{\mu\nu}\equiv g^{\mu\nu}\left(\Gamma^\alpha_{\mu\nu,\alpha}-\Gamma^\alpha_{\mu\alpha,\nu}+\Gamma^\alpha_{\alpha\lambda}\Gamma^\lambda_{\mu\nu}-\Gamma^\alpha_{\mu\lambda}\Gamma^\lambda_{\alpha\nu}\right),\nonumber\\
		\Re= g^{\mu\nu}\Re_{\mu\nu}\equiv g^{\mu\nu}\left(\hat{\Gamma}^\alpha_{\mu\nu,\alpha}-\hat{\Gamma}^\alpha_{\mu\alpha,\nu}+\hat{\Gamma}^\alpha_{\alpha\lambda}\hat{\Gamma}^\lambda_{\mu\nu}-\hat{\Gamma}^\alpha_{\mu\lambda}\hat{\Gamma}^\lambda_{\alpha\nu}\right).
	\end{eqnarray}
 
However, in the metric approach, curvature depends only on the metric, while in the Palatini approach curvature is a function of both the metric and the affine connection.
Like other  metric and Palatini  models, it is possible to represent HMPG in a scalar-tensor form. Let us to introduce an  auxiliary field $A$ that allows to write the action as follows:
	\begin{equation}
		\label{action2}
		S=\frac{1}{2k^2}\int d^4x\sqrt{-g}[R+f(A)+f_A\cdot(\Re-A)]+S_m,
	\end{equation}
where $f_A=\frac{df}{dA}$. Then we can introduce the scalar field $\phi\equiv f_A$ and the potential $V(\phi)\equiv Af_A-f(A)$, which consist of kinetic and potential parts. After mathematical transformations  we obtain the scalar-tensor representation of the HMPG action:
	\begin{equation}
		\label{action3}
		S=\frac{1}{2k^2}\int d^4x\sqrt{-g}[R+\phi\Re-V(\phi)]+S_m.
	\end{equation}

To obtain the field equations, it is necessary to vary the action with respect to all variables: the metric tensor $g_{\mu\nu}$, the scalar field $\phi$ and the affine connection $\hat{\Gamma}^\alpha_{\mu\nu}$. Thus the field equations take the following form: 
	\begin{eqnarray}    
	\label{fieldeq}
		R_{\mu\nu}+\phi\Re_{\mu\nu}-\frac{1}{2}\left(R+\phi\Re-V\right)g_{\mu\nu}&=&k^2T_{\mu\nu},\nonumber\\
		\Re-V_\phi&=&0,\\\nonumber
		\hat{\nabla}_\alpha\left(\sqrt{-g}\phi g^{\mu\nu}\right) &=& 0,
	\end{eqnarray}
where $T_{\mu\nu}$ is the energy-momentum tensor, and $V_\phi\equiv\frac{dV}{d\phi}$. It is necessary to emphasize that there is no torsion in HMPG. Consequently, the final equation can be resolved, revealing that the independent connection corresponds to the Levi-Civita connection of the metric $h_{\mu\nu} = \phi g_{\mu\nu}$. By employing the solution of this equation, a connection between the metric and the Palatini curvatures can be expressed:
 	\begin{equation}
		\label{R(Re)}
		\Re_{\mu\nu}=R_{\mu\nu}+\frac{3}{2\phi^2}\partial_\mu\phi\partial_\nu\phi-\frac{1}{\phi}\left(\nabla_\mu\nabla_\nu\phi+\frac{1}{2}g_{\mu\nu}\nabla_\alpha\nabla^\alpha\phi\right).
	\end{equation}
This connection allows to rewrite the action without Palatini terms \cite{Harko2012, Capozziello2015}:
	\begin{equation}
		\label{action4}
		S=\frac{1}{2k^2}\int d^4x\sqrt{-g}\left[(1+\phi) R+\frac{3}{2\phi}\partial_\mu\phi\partial^\mu\phi-V(\phi)\right]+S_m.
	\end{equation}
Thus, the action of HMPG in the scalar-tensor representation has been obtained.

Taking into accaunt expressions~(\ref{fieldeq}), it is possible to derive the scalar-tensor form of the field equations:
	\begin{eqnarray}
		\label{feq_g}
		\frac{1}{1+\phi}\Bigg[k^2\left(T_{\mu\nu}-\frac{1}{2}g_{\mu\nu}T\right)&+&\frac{1}{2}g_{\mu\nu}\left(V+\nabla_\alpha\nabla^\alpha\phi\right)+\nabla_\mu\nabla_\nu\phi \nonumber\\
&-&\frac{3}{2\phi}\partial_\mu\phi\partial_\nu\phi\Bigg]=R_{\mu\nu},
\end{eqnarray}
\begin{equation}
		\label{feq_phi}
		-\nabla_\mu\nabla^\mu\phi+\frac{1}{2\phi}\partial_\mu\phi\partial^\mu\phi+\frac{\phi[2V-(1+\phi)V_\phi]}{3}=\frac{\phi k^2}{3}T.
	\end{equation}
 
From here onwards,  the field equations are considered in this form.
\section{ Thin accretion disk model}\label{sec4}
	
An accretion disk is an astrophysical  structure  that forms near a massive object, representing diffuse material orbiting around the central body. In this article only thin accretion discs are considered. The general model of such disks was developed by N. Shakura and R. Sunyaev \cite{Shakura1973} and later extended by I. Novikov, K. Thorne and D. Page \cite{Page1974,Novikov1973}. A thin accretion disc is characterized by the fact that its vertical size, $h$, is negligible compared to its horizontal size, $h<< r$. In such structures, particles move along Keplerian orbits, and the accretion disk is located in the equatorial plan of the compact body. Moreover, the thin accretion disc model  implies  a steady state. Thus, the accretion mass rate, $\dot M_0$, is assumed to be constant over time. Additionally, in the steady-state model the accreting matter is  in thermodynamical equilibrium \cite{mohaddese}. 

To study the electromagnetic properties of an accretion disk, it is first necessary to study the geometry of space in which particles move near a compact object \cite{Harko2011}. The geodesic motion of test particles moving around a massive body is governed by the Lagrangian:
\begin{equation}\label{lagrangian}
\mathcal{L}=\frac{1}{2}g_{\mu\nu}\dot x^\mu \dot x^\nu,
\end{equation}
where the dot means the derivative with respect to $\tau$, which is an affine parameter along the
geodesic $x^\mu(\tau)$; $g_{\mu\nu}$ is the metric of a static, spherically symmetric space-time:
	\begin{equation}
	ds^2=g_{00}dt^2+g_{11}dr^2+g_{22}d\theta^2+g_{33}d\varphi^2,
	\end{equation}
	for which the elements $g_{00}, g_{11}, g_{22}, g_{33}$ only depend on the radial coordinate $r$. We also use an equatorial approximation, implying that $|\theta - \pi/2| << 1$. To derive the main features of the thin accretion disc, it becomes essential to define the specific energy $\tilde E$ and the specific angular momentum $\tilde L$ using the Euler-Lagrange equations \cite{Page1974}:
		\begin{equation} \label{E}
		g_{00}\dot t=\tilde E,
	\end{equation}	
		\begin{equation}\label{L}
		g_{22}\dot \varphi=\tilde L.
	\end{equation}
 	Here $\tilde E=E/m_0c^2$ and $\tilde L=L/m_0c$, where $E$ represents the total energy of the particle in its orbit, $m_0c^2$ is the rest mass energy of this particle and $L$ is the angular momentum of the particle.

Another important characteristic of the accretion disc is the effective potential $V_{eff}(r)$. It can be defined from the relation $2\mathcal{L} =-1$, taking into accaunt equations (\ref{E}) and (\ref{L}):
			\begin{equation}
			-g_{00}g_{11}\dot r^2+V_{eff}(r)=\tilde E^2.
	\end{equation}
	Then the effective potential is defined as
		\begin{equation}
		V_{eff}(r)=-g_{00}\biggl(1+\frac{\tilde L^2}{g_{33}}\biggr).
	\end{equation}
	The circular orbit passes where the minimum of the effective potential is located. If there is no minimum (i.e., the potential has a "smooth" shape), then the circular orbits are unstable for a given moment. Therefore, for stable circular orbits, conditions $V_{eff}(r) = 0$ and $V_{eff,r}(r) = 0$ must be satisfied. This makes it possible to determine the specific energy, specific angular momentum, and angular velocity $\Omega$ of particles moving within the gravitational potential of a massive object:
		\begin{equation}\label{E1}
		\tilde E=-\frac{g_{00}}{\sqrt{-g_{00}-g_{33}\Omega^2}},
	\end{equation}	
		\begin{equation}\label{L1}
		\tilde L=\frac{g_{33}\Omega}{\sqrt{-g_{00}-g_{33}\Omega^2}},
	\end{equation}	
		\begin{equation}
		\Omega=\frac{d\varphi}{dt}=\sqrt{\frac{-g_{00,r}}{-g_{33,r}}}.
	\end{equation}
	The condition $V_{eff,rr}(r) = 0$ determines the radius of  the innermost stable circular orbit $r_{isco}$. This leads to the following equation:

		\begin{equation}
		\tilde E^2 g_{33,rr}+\tilde L^2 g_{00,rr}+(g_{00}g_{33})_{,rr}=0.
	\end{equation}
	Solving this equation with respect to $r$ and taking into account expressions (\ref{E1}) and (\ref{L1}), we obtain the $r_{isco}$.

A key feature of an accretion disk is its accretion efficiency, a measure which signifies the capacity of the central body to transmute rest mass into outgoing radiation. This measure is established as the proportion of the photon energy rate that escapes from the disk's surface to infinity to the rate at which mass-energy is transported to  the black hole. If all emitted photons can escape to infinity, the efficiency can be expressed in terms of the specific energy defined at $r_{isco}$ \cite{Page1974,Novikov1973}:

		\begin{equation}
		\eta=1-\tilde E_{ms}.
	\end{equation}
		Schwarzschild black holes  have an efficiency of approximately 6\%, while for extreme rotating Kerr black holes, $\eta \approx 42\%$. It is important to note that the photon capture by the black hole can influence the efficiency. For instance, a Kerr black hole with photon capture has an efficiency of 40\% \cite{kipthorne}.

		One of the main parameters that characterize the spectrum of an accretion disk is the time averaged energy flux emitted from the surface of the disk. It can be obtained from the conservation equations $\nabla_\mu E^\mu=0$ and $\nabla_\mu J^\mu=0$ \cite{Page1974}. These laws follow from the law of conservation of the energy-momentum tensor. Initially, in the work of D. Page and K. Thorne \cite{Page1974}, the energy flux was obtained within the framework of GR, where, as is known, the energy-momentum tensor is locally conserved. In our work we consider another theory, HMPG. In this theory, local conservation laws are also satisfied, as is shown in the work \cite{Tian_2016}. The authors investigate the problem of conservation laws in scalar-tensor theories of gravity and theories that have a scalar-tensor representation. It is shown that HMPG has divergence-free field equations and respect the local energy-mentum conservation. Therefore, the general form of the formula for energy flux is also applicable in this model, as in GR. The radiation flux per unit area can be expressed in terms of the specific energy, angular momentum and the angular velocity of the particles orbiting in the disk as follows:
		\begin{equation}\label{page_thorne_flux}
		F(r)=-\frac{\dot M_0}{4\pi\sqrt{-g}}\frac{\Omega_{,r}}{(\tilde E-\Omega \tilde L)^2} \int^r_{r_{isco}}(\tilde E-\Omega \tilde L) \tilde L_{,r}rdr,
	\end{equation}
	where $\dot M_0$ is the mass accretion rate. This quantity measures the rate at which the rest mass of the particles streams inward through the disk with respect to the coordinate time.~\footnote{In the original article by D. Page and K. Thorne \cite{Page1974}, this formula (\ref{page_thorne_flux}) was obtained for cylindrical coordinates. Therefore, when working in a spherical coordinate system, it is important to take into account the Jacobian for the transition from one system to another.}

The steady-state thin disk model includes a condition that the accreting matter is assumed to be in thermodynamic equilibrium. As a result, an ideal black body model is applicable in describing the radiation from the disk's surface, and the energy flux can be derived from the Stefan-Boltzmann law, $F(r) = \sigma T(r)^4$, where $\sigma$ is the Stefan-Boltzmann constant. Therefore, the observed luminosity has a redshifted black body spectrum as follows \cite{torres}:
		\begin{equation}
		L(\nu)=\frac{2h}{ c^2}\cos{\gamma}\int^{r_f}_{r_i}\int^{2\pi}_0\frac{\nu_e^3rd\varphi dr}{\exp{(h\nu_e/kT)}-1}.
	\end{equation}
	Here $d$ denotes the distance to the source, $\gamma$ represents the disk inclination angle, $r_i$ and $r_f$ are the radii of the disk's inner and outer edge, respectively; $\nu_e = \nu(1 + z)$ denotes the emitted frequency, and the redshift factor is defined as:
		\begin{equation}
		1+z=\frac{1+\Omega r \sin{\varphi}\sin{\gamma}}{\sqrt{-g_{00}-g_{33}\Omega^2}},
	\end{equation}
	where we neglect the light bending \cite{luminet}.
  \section{Numerical results}\label{sec4}
 The spherically symmetric space-time outside massive objects can be described by the following line element:
 	\begin{equation} \label{metric}
 	ds^2=-e^{\nu(r)}dt^2+e^{\lambda(r)}dr^2+r^2(d\theta^2+\sin^2\theta d\varphi^2).
 	\end{equation}
 	 The metric functions $\nu(r)$ and $\lambda(r)$ depend only on the radial coordinate $r$, with the condition $0 \leq r < \infty$.  Taking into account the metric (\ref{metric}), the field equations (\ref{feq_g}) and (\ref{feq_phi}) can be represented in the following form \cite{Danila2019}:
 \begin{equation}   
 \frac{d\phi}{d\xi}=-\frac{U}{\xi^2},
 \end{equation}
  \begin{eqnarray}
 \frac{dM_{eff}}{d\xi}=&\frac{(1-M_{eff}\xi)[\xi^2dU/d\xi+3U^2/4\phi-2\xi U]+M_{eff}\xi^3(1+\phi)-v}{\xi^4(1+\phi+U/2\xi)} \nonumber\\
&-\frac{M_{eff}}{\xi},
 \end{eqnarray}

 \begin{equation}
  \frac{d\nu}{d\xi}=-\frac{\xi-\bigl\{\frac{U(\xi)[8\phi+3U(\xi)/\xi]}{4\phi(1+\phi)}+\xi\bigr\}[1-\xi M_{eff}(\xi)]-\frac{v(\phi)}{\xi(1+\phi)}}{\xi^2[1-\xi M_{eff}(\xi)]\bigl[1+\frac{U(\xi)}{2\xi(1+\phi)}\bigr]},
 \end{equation}
\begin{align}
\frac{d^2\nu}{d\xi^2}=&\frac{(1-\frac{\xi}{2}\frac{d\nu}{d\xi})(-\xi\frac{dM_{eff}}{d\xi}-M_{eff})}{\xi(1-\xi M_{eff})}-\frac{5U(\xi)^2}{2\xi^4\phi(1+\phi)}\nonumber\\
 &-\frac{2}{\xi^4(1+\phi)(1-\xi M_{eff})}\biggl\{\frac{2\phi}{3}[2v-(1+\phi)v_\phi]+v\biggr\}\nonumber\\
&+\frac{2U}{\xi^3(1+\phi)}-\frac{1}{2}\biggl(\frac{d\nu}{d\xi}\biggr)^2+\frac{1}{\xi}\frac{d\nu}{d\xi}, 
\end{align}
\begin{gather}
\begin{split}
  \frac{dU(\xi)}{d\xi}=&\frac{\frac{\xi^2 U(\xi)}{2}\bigl[\xi\frac{d M_{eff}(\xi)}{d\xi}+M_{eff}(\xi)\bigr]}{\xi^2(1-\xi M_{eff}(\xi))}+\frac{2U(\xi)}{\xi}-\frac{1}{\xi^2}\frac{U^2(\xi)}{2\phi}\\
&-\frac{\frac{2\phi}{3}\bigl[2v(\phi)-(1+\phi)v_\phi(\phi)\bigr]}{\xi^2(1-\xi M_{eff}(\xi))}-\frac{U(\xi)}{2}  \frac{d\nu}{d\xi}. 
\end{split}
\end{gather}

To obtain these equations, the following transitions to dimensionless variables were used
\begin{equation}
\xi=\frac{2GM_{BH}}{c^2r},\ \ \ \ \frac{d\phi}{d r}=\frac{c^2}{2GM_{BH}}U, \ \ \ \ 
V(\phi)=2\biggl(\frac{c^2}{2GM_{BH}}\biggr)^2v(\phi),
\end{equation}
where $M_{BH}$ is the black hole mass.
The metric function $e^{-\lambda(r)}$ is redefined as
\begin{equation}
e^{-\lambda(r)}=1-\frac{2GM_{BH} M_{eff}(r)}{c^2r}.
\end{equation}

These field equations was obtained in the article \cite{Danila2019}. Then authors found numerical solutions for these field equations. In our work, we repeat the numerical analysis using integration methods from Python's scipy library. For the solution, the fixed initial conditions
\begin{equation}
M_{eff} (0) = 1, \ \ \ \nu(0) = 0, \ \ \ \nu'(0) = 0
\end{equation}
were taken into account,
and arbitrary numerical values for 
\begin{equation}
u(0) = u_0, \ \ \ \  \phi(0) =\phi_0
\end{equation} 
were used.

The details of the calculations of the metric can be found in the article \cite{Danila2019}.
\section{Properties of thin accretion discs}
Now, we are ready to investigate the properties of thin accretion disks around black holes in hybrid metric-Palatini gravity and compare the results with  the Schwarzschild predictions. We consider two cases: a case without potential $V=0$ and a case where potential takes the Higgs-type form $V=-\frac{\mu^2}{2}\phi^2+\frac{\zeta}{4}\phi^4$.

It is important to emphasize that in this article we consider the case of a static spherically symmetric black hole. To obtain the energy flux, emission spectrum, temperature, and efficiency, we use the actual values of mass and accretion rate of the system MAXI J1820+070 \cite{Zhao2021}. This system is chosen due to the small value of the Kerr parameter $a=0.14$, which is closest to a Schwarzschild black hole.
\subsection{Case $V=0$}
In the case $V=0$, the metric of the HMPG includes two model parameters: the initial value of the scalar field, $\phi_0$, and the initial value of its derivative $u_0$. We consider three cases:
\begin{enumerate}
\item fixed $\phi_0=1$ and range of $u_0=[4\times 10^{-9}; 6.4\times 10^{-8}]$,
\item fixed $u_0=5.12\times 10^{-7}$ and range $\phi_0=[0.5;8]$,
\item a connection between $\phi_0$ and $u_0$ obtained from the post-Newtonian analysis. This connection has the following form:
\begin{equation}\label{u0}
u_0=\frac{2GM\phi_0}{3c^2r^2}.
\end{equation} 
It can be obtained from the expression for scalar perturbation $\varphi=\frac{-2GM\phi_0e^{-m_\phi r}}{3c^2r}$ \cite{Dyadina2019}. In this case, we consider $\phi_0<4\times10^{-5}$ \cite{Leanizbarrutia2017}. This limit was imposed on the initial value of the scalar field using the data from the Cassini experiment \cite{Cassini2003}.
\end{enumerate}

The first two cases are chosen due to the fact that such black holes were studied in the article \cite{Danila2019}, where spherically symmetric solution was obtained. The choice of the last case has the following reason: we take the initial value of the scalar field  at a sufficiently large distance from the black hole. At this distance, the gravitational field is quite weak, therefore we can use the results of the post-Newtonian analysis.

Further we present the results of the analysis of figures in accordance with these cases.
\subsubsection{Energy flux}
\begin{figure*}
		\begin{minipage}[h]{.45\textwidth}
			\includegraphics[width=\columnwidth]{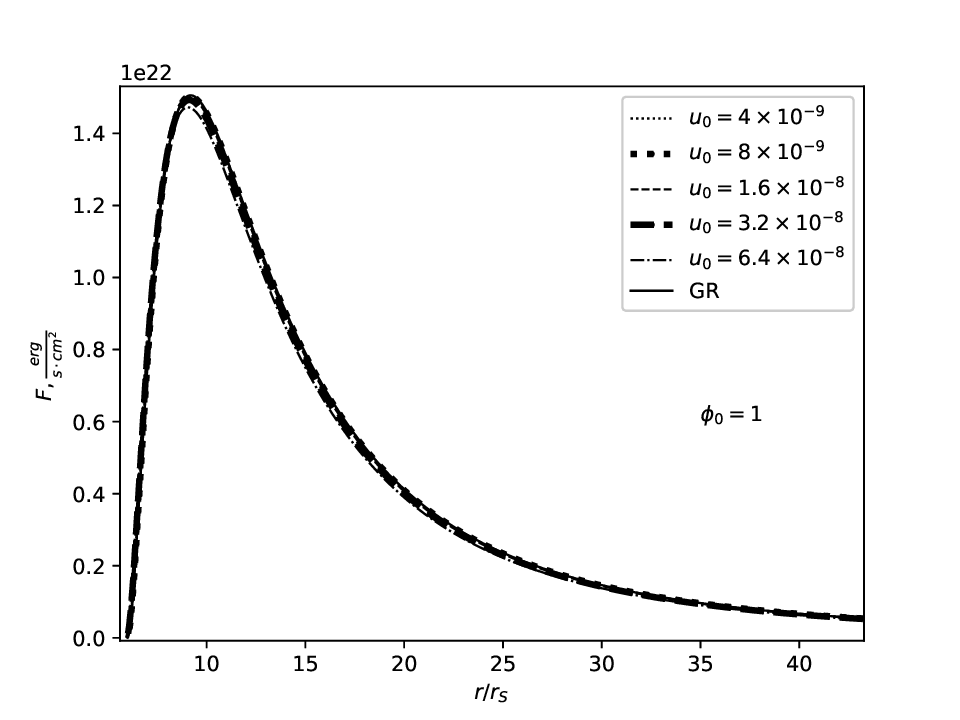} \\ a)
					\end{minipage}
\hfill
		\begin{minipage}[h]{.45\textwidth}
			{\includegraphics[width=\columnwidth]{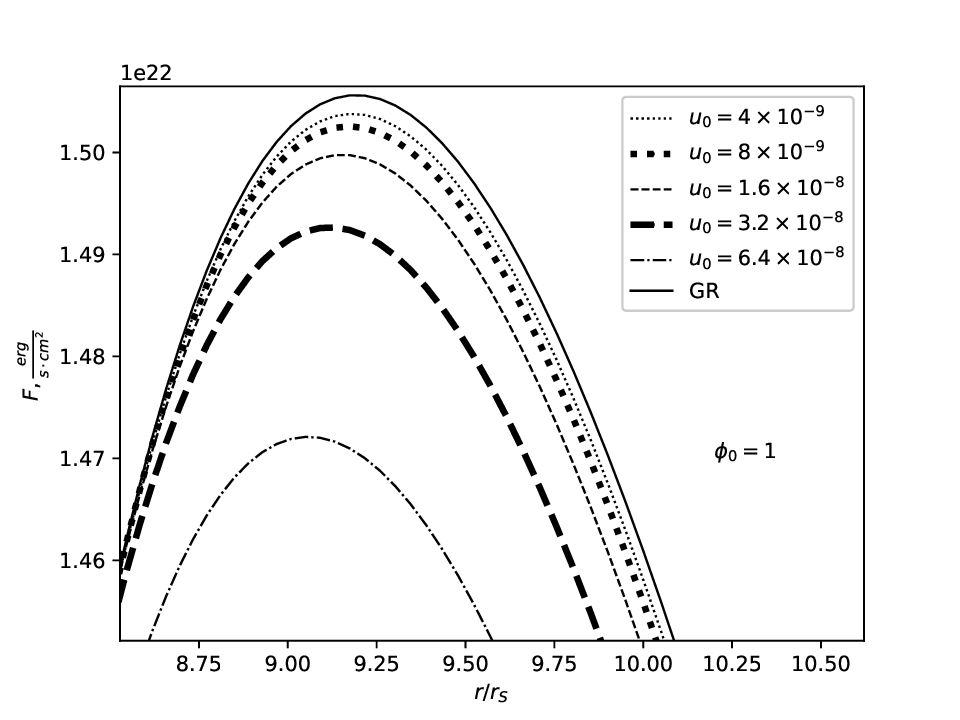}} \\ b)
			
		\end{minipage}

		\hfill

		\begin{minipage}[h]{.45\textwidth}
			\includegraphics[width=\columnwidth]{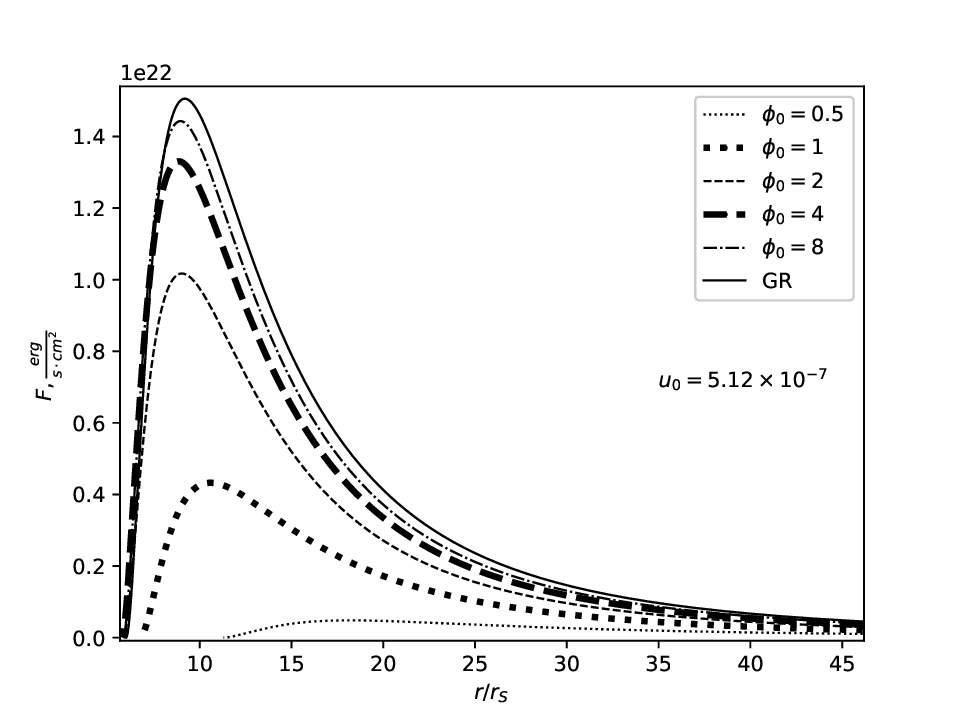} \\ c)

	\end{minipage}
\hfill
		\begin{minipage}[h]{.45\textwidth}
			\includegraphics[width=\columnwidth]{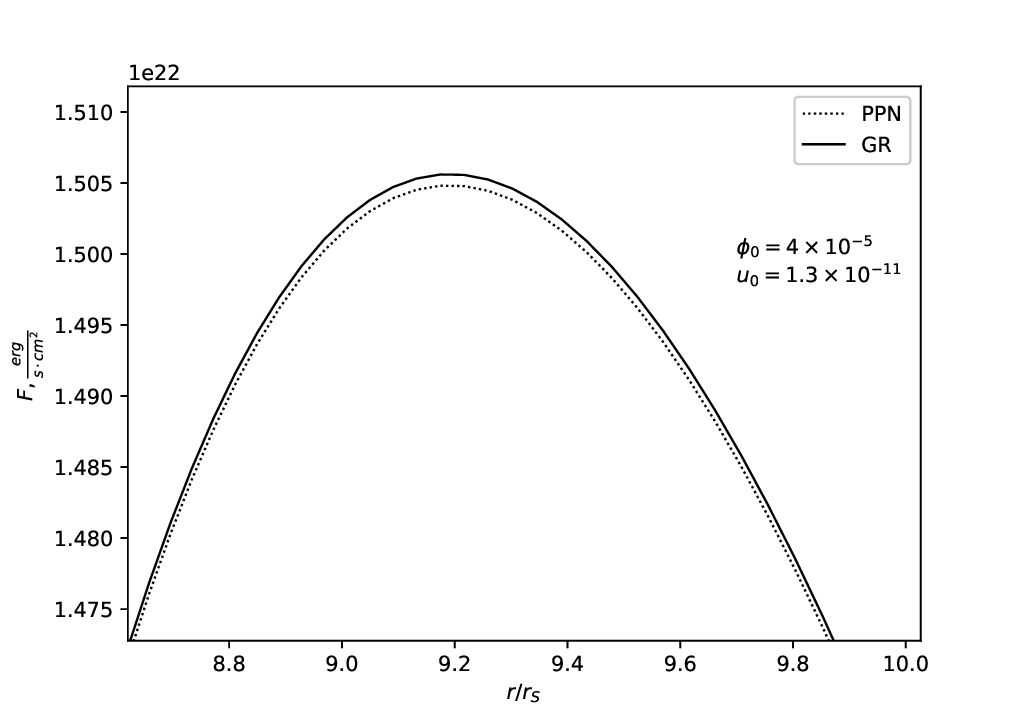} d)

\end{minipage}
		\caption{Case $V=0$.  The energy flux $F(r)$ of a disk around a static black hole with  $\dot M = 2.21\times10^{18} g/s$ and $M=8.48 M_\odot$   as function of the normalized radial coordinate $r/r_s$. b) is zoom version of a). In d) the connection \eqref{u0} from PN analysis is taking into accaunt.	}
		
	\label{fig:flux1}
	\end{figure*}

 \begin{enumerate}
\item The effects of HMPG on the energy flux across the disk's surface are presented in Fig. \ref{fig:flux1}a, \ref{fig:flux1}b in this case. Here, all curves lie below the Schwarzschild solution. Additionally, as $u_0$ increases, the peak value of the energy flux decreases. The largest discrepancy in the peak energy flux within this range of parameters is no more than $2.5 \%$ relative to the Schwarzschild value.
\item In Fig. \ref{fig:flux1}c, we observe a decline in the maximum energy flux as $\phi_0$ decreases. All curves again sit beneath the Schwarzschild one. However, in this case, the difference between the maxima is much more noticeable, reaching up to 95\%. 
\item The most realistic picture arises when the parameters are considered within the limits imposed from the solar system.  As we can see in the Fig. \ref{fig:flux1}d in this case the result is close to Schwarzschild curve.
\end{enumerate}

\subsubsection{Emission spectra.}
\begin{figure*}
		\begin{minipage}[h]{.45\textwidth}
			\includegraphics[width=\columnwidth]{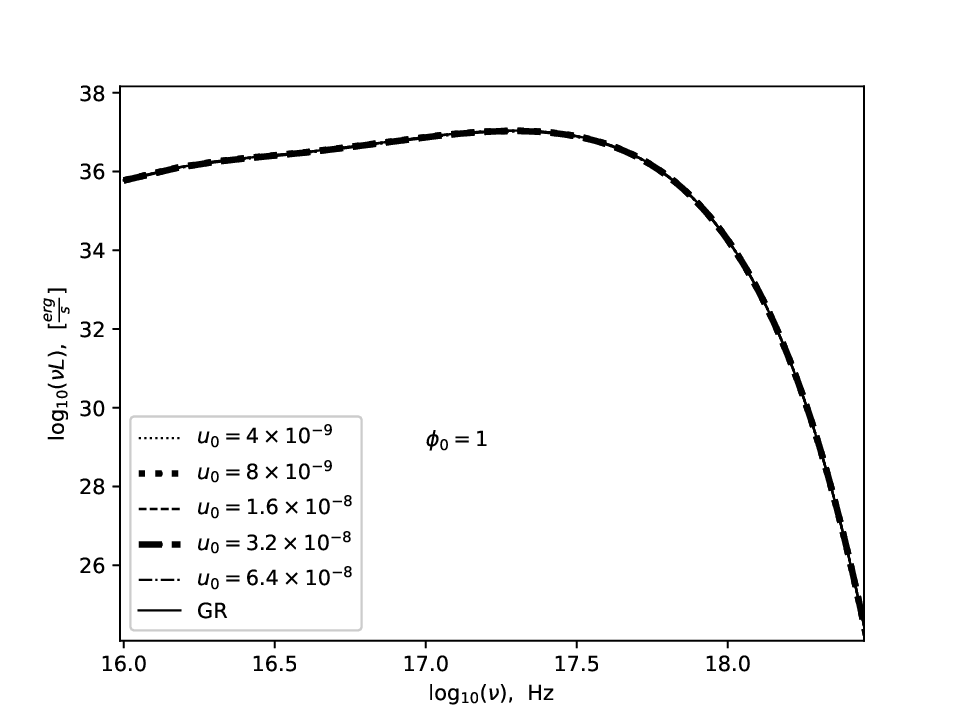} \\ a)
					\end{minipage}
\hfill
		\begin{minipage}[h]{.45\textwidth}
			{\includegraphics[width=\columnwidth]{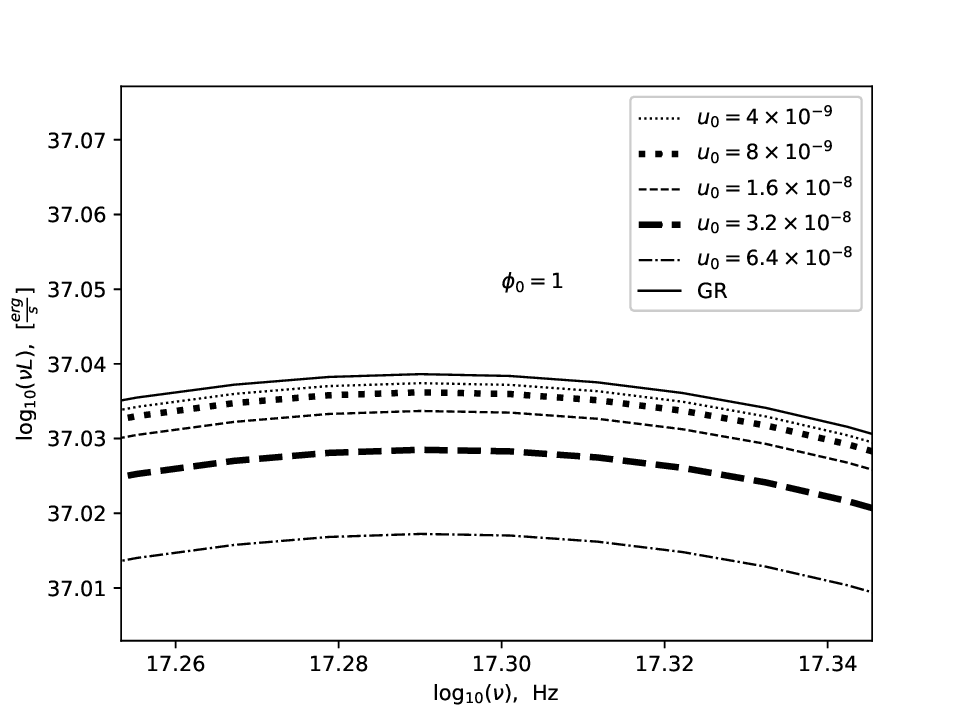}} \\ b)
			
		\end{minipage}

	\hfill
		\begin{minipage}[h]{.45\textwidth}
			\includegraphics[width=\columnwidth]{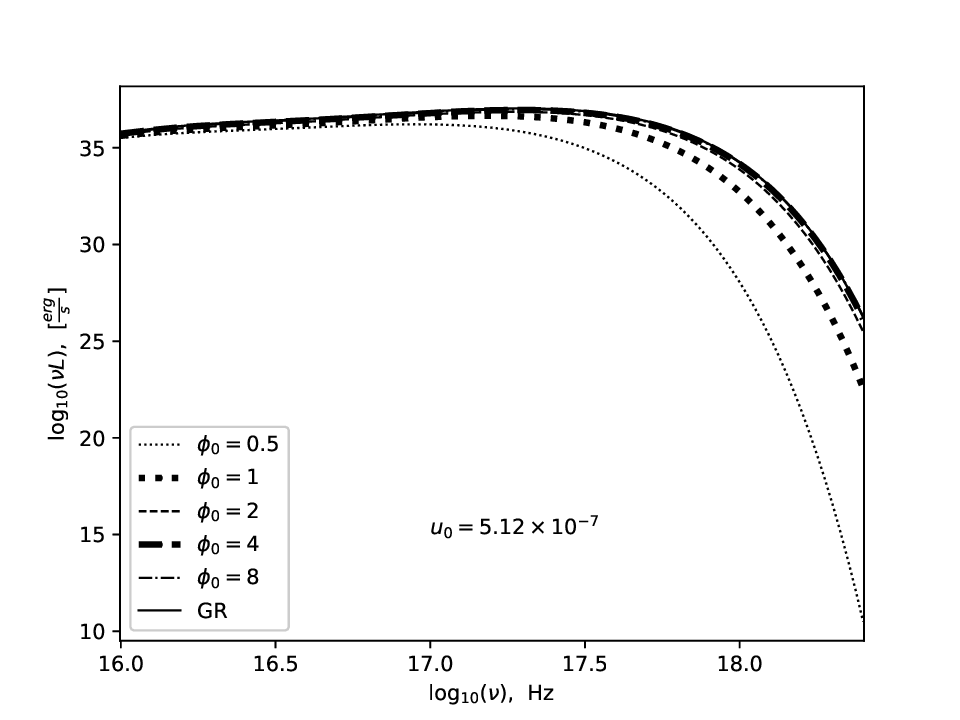} \\ c)
					\end{minipage}
\hfill
		\begin{minipage}[h]{.45\textwidth}
			{\includegraphics[width=\columnwidth]{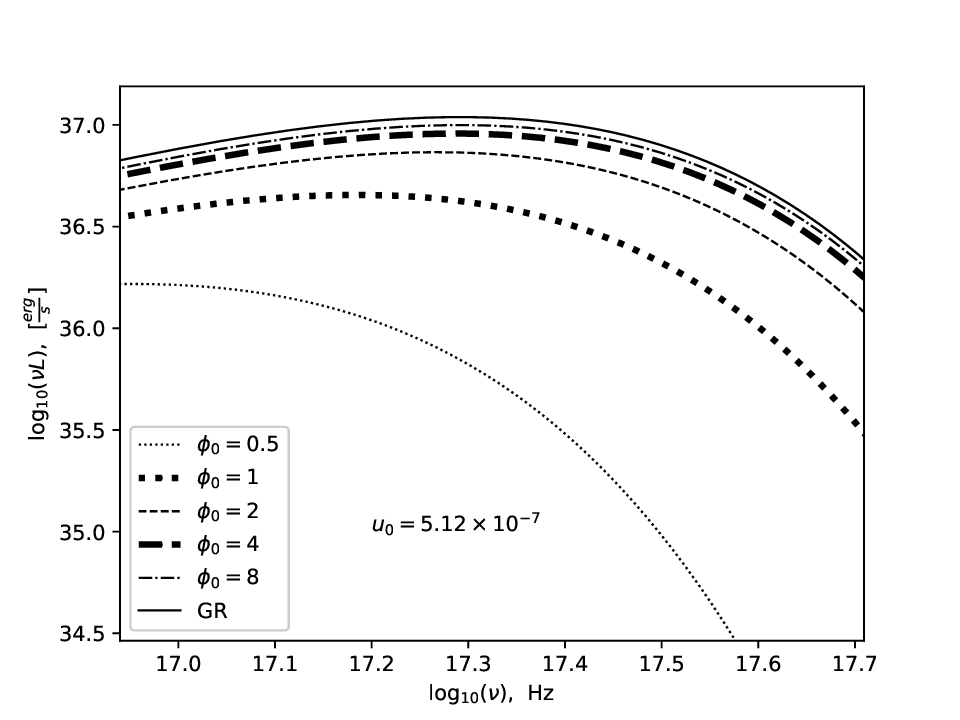}} \\ d)
		
		\end{minipage}
		
		\hfill

\begin{center}
		\begin{minipage}[h]{.55\textwidth}
			\includegraphics[width=\columnwidth]{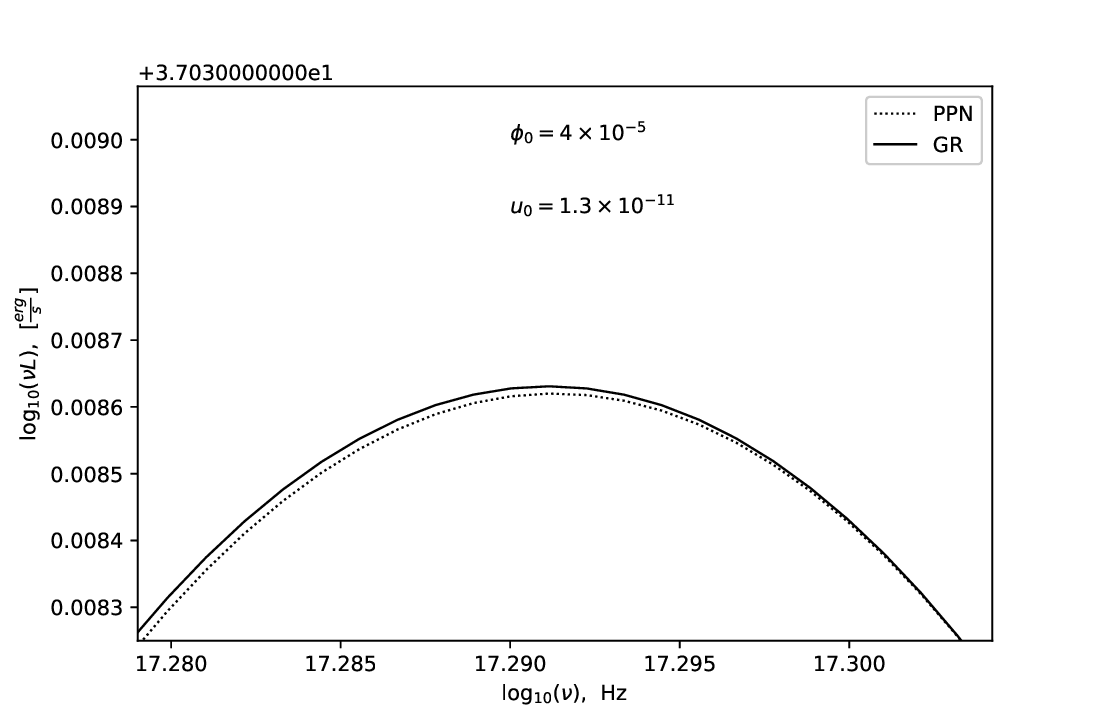} \\ e)

	\end{minipage}
	\end{center}
			\caption{Case $V=0$. The emission spectrum $\nu L(\nu)$ of the accretion disk 	around a static black hole with $\dot M = 2.21\times10^{18} g/s$ and $M=8.48 M_\odot$  as function of frequency $\nu$. b) is zoom version a); d) is zoom version c).	In e) the connection \eqref{u0} from PN analysis is taking into accaunt.}
\label{fig:lum1}
	\end{figure*}

\begin{enumerate}
\item In Fig. \ref{fig:lum1}a,  \ref{fig:lum1}b we display the emission spectra for accretion disk around black holes and compare it with the emission spectra for the Schwarzschild black hole.  In the case of a fixed $\phi_0$, the luminosity decreases with increasing $u_0$. 
\item For a fixed $u_0$, the luminosity decreases with decreasing $\phi_0$. It can be seen on Fig. \ref{fig:lum1}c, \ref{fig:lum1}d. 
\item The curve obtained in the frameworks of solar system limitations  looks the most realistic. The two curves are almost identical in Fig. \ref{fig:lum1}e.
\end{enumerate}

\subsubsection{Temperature.}
\begin{figure*}
		\begin{minipage}[h]{.45\textwidth}
			\includegraphics[width=\columnwidth]{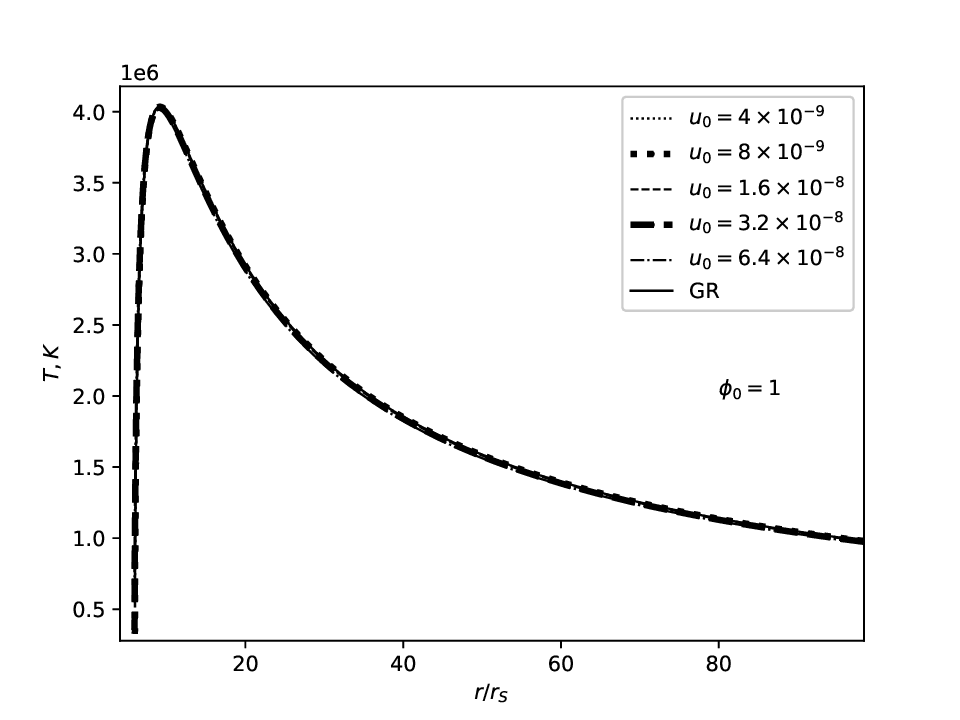} \\ a)
					\end{minipage}
\hfill
		\begin{minipage}[h]{.45\textwidth}
			{\includegraphics[width=\columnwidth]{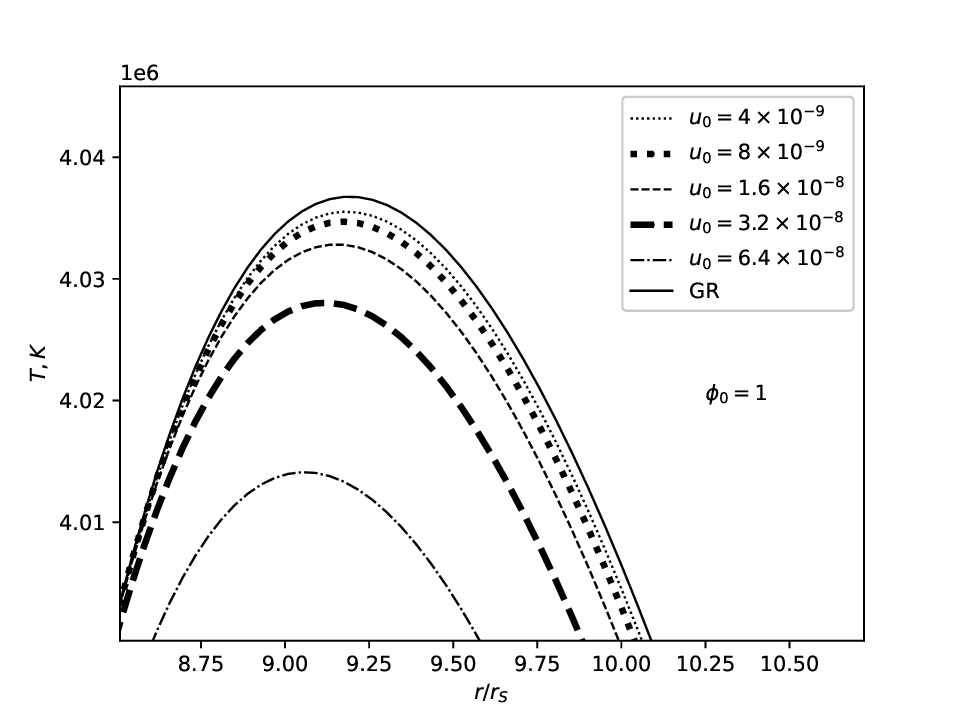}} \\ b)
		
		\end{minipage}

\hfill
		\begin{minipage}[h]{.45\textwidth}
			\includegraphics[width=\columnwidth]{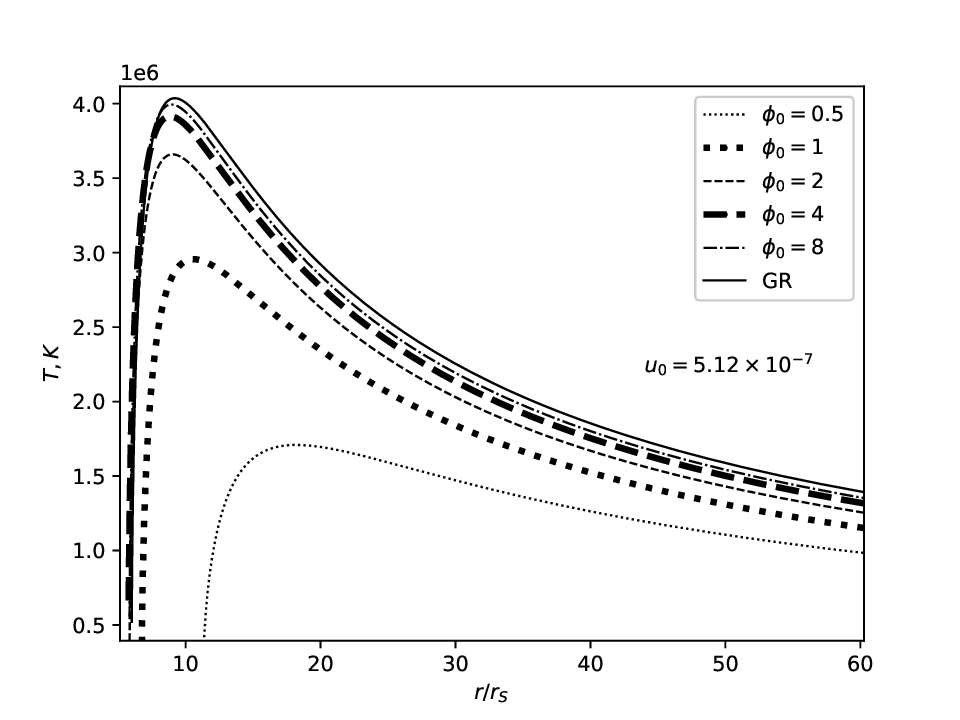} \\ c)

	\end{minipage}
\hfill
		\begin{minipage}[h]{.45\textwidth}
			\includegraphics[width=\columnwidth]{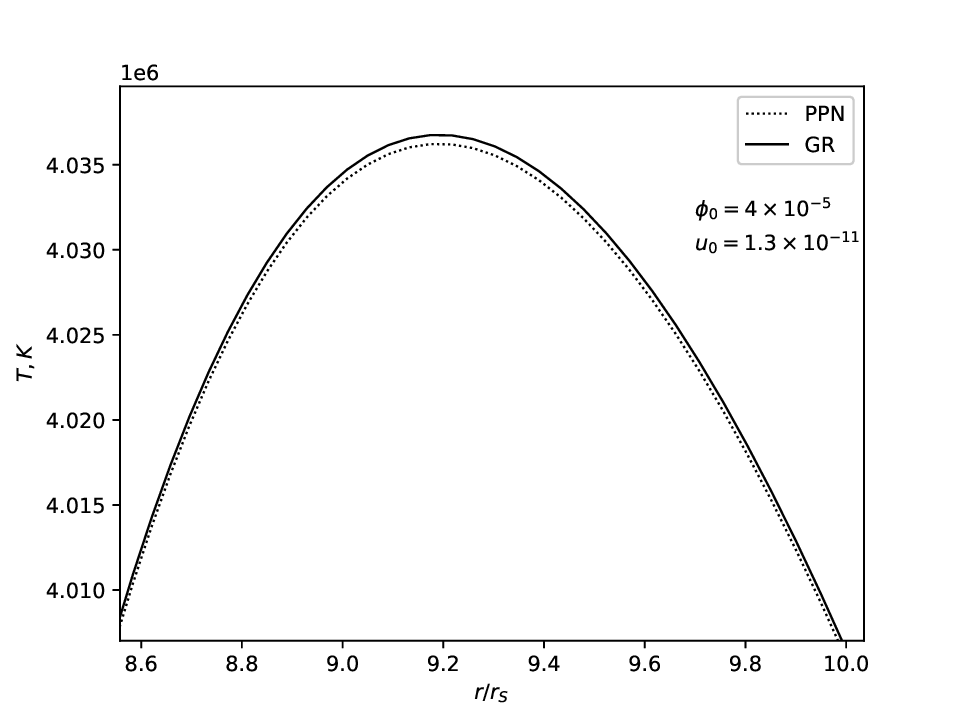} \\ d)	
\end{minipage}
		\caption{Case $V=0$.  The temperature distribution $T(r)$ of a disk around a static black hole with $\dot M = 2.21\times10^{18} g/s$ and $M=8.48 M_\odot$ as function of the normalized radial coordinate $r/r_s$. b) is zoom version a).	 In d) the connection \eqref{u0}  from PN analysis is taking into accaunt.}
		
\label{fig:T1}

	\end{figure*}

Since the temperature is related to the energy flux through the Stefan-Boltzmann constant, the results for the effective temperature are no different from the conclusions made for the energy flux. These results are illustrated in Fig. \ref{fig:T1}.

Generally speaking, temperature can serve as an excellent indicator to test a theory because its value can be determined from observational data \cite{Zhao2021}. Unfortunately, all known accreting systems have a nonzero Kerr parameter. And even a small deviation of this parameter from zero (for example, 0.14) does not allow us to check the results obtained for a static spherically symmetric black hole with observational data.  Thus, the observational test of the theory is possible only in the case of obtaining the Kerr-type solution and the accretion characteristics for such a black hole.

\subsubsection{Efficiency.}
\begin{figure*}
		\begin{minipage}[h]{.3\textwidth}
			\includegraphics[width=\columnwidth]{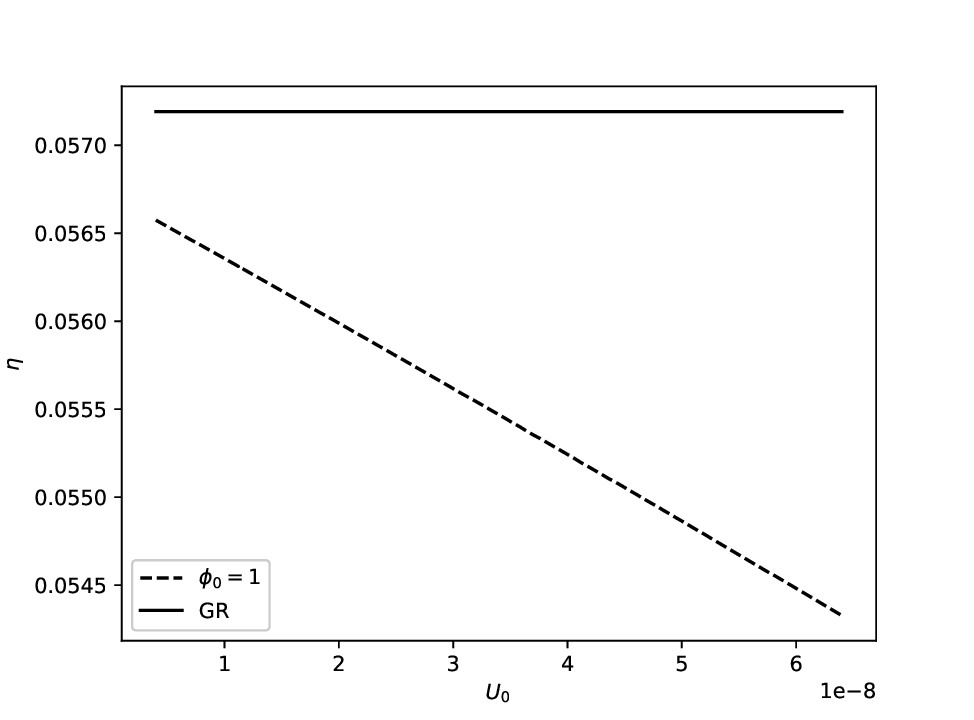} \\ a)
				
					\end{minipage}
\hfill
		\begin{minipage}[h]{.3\textwidth}
			{\includegraphics[width=\columnwidth]{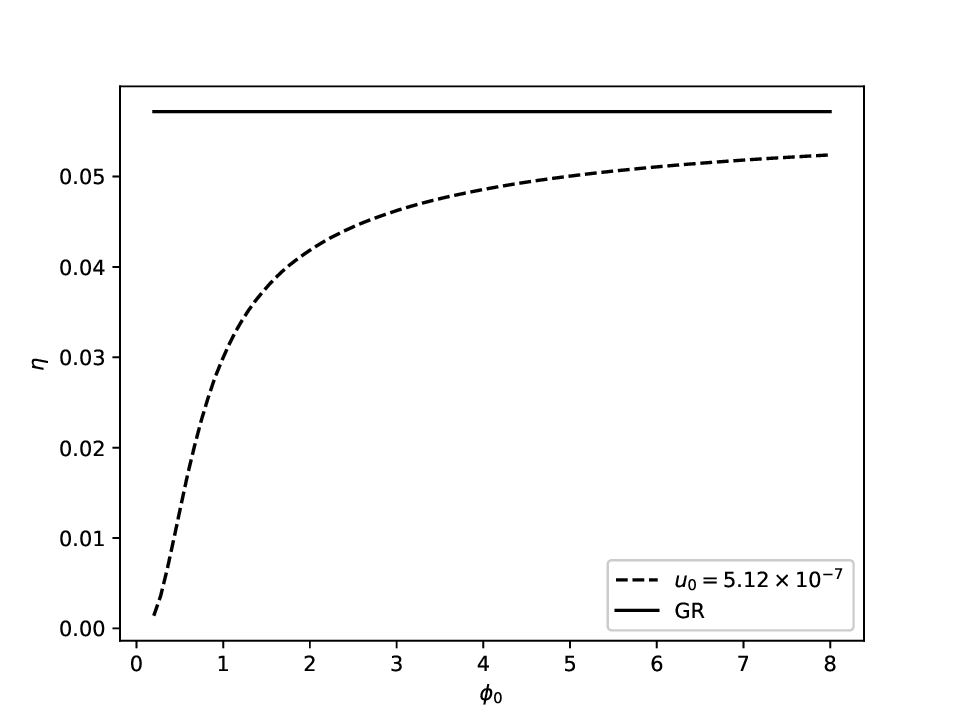}} \\ b)

		\end{minipage}
\hfill 
\begin{minipage}[h]{.3\textwidth}
			{\includegraphics[width=\columnwidth]{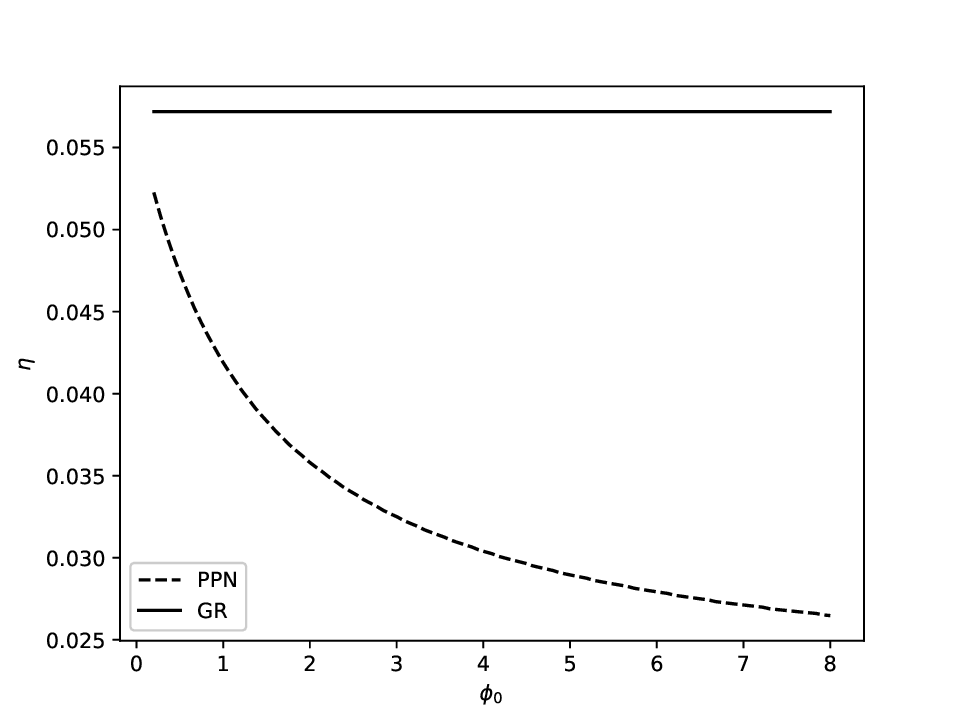}} \\ c)
			
			\end{minipage}
			\caption{Case $V=0$.  The efficiency for thin accretion disk around static black hole  a)  as function of  $u_0$; 	 b)   as function of  $\phi_0$;  c)  as function of  $\phi_0$, taking into accaunt connection (\ref{u0}) between $\phi_0$ and $u_0$ (from PN analysis). 		}

		\label{fig:eff}
	
	\end{figure*}

The discussion of  efficiency deserves special attention.
\begin{enumerate}
\item The case of fixed $\phi_0$ is shown in Fig. \ref{fig:eff}a. As $u_0$ increases, the efficiency decreases.  
\item with the growth of $\phi_0$ the efficiency increases and its value approaches the Schwarzschild one. Moreover, small values of $\phi_0<1$ may differ by more than 80\% from the predictions of GR (Fig. \ref{fig:eff}b). 

\item If we take into account the relationship between $u_0$ and $\phi_0$ (\ref{u0}), we can see that as $\phi_0$ increases, the efficiency decreases. This dependence is illustrated in  Fig. \ref{fig:eff}c. The difference with the maximum value is almost 50\%. On the other hand, we can observe that no value of $\phi_0$ achieves the efficiency value for a Schwarzschild black hole. 
\end{enumerate}

\subsection{Case $V=-\frac{\mu^2}{2}\phi^2+\frac{\zeta}{4}\phi^4$}

In the article \cite{Danila2019} authors consider the only one case with the potential, and this potential has a Higgs-type form:
\begin{equation}
V=-\frac{\mu^2}{2}\phi^2+\frac{\zeta}{4}\phi^4,
\end{equation}
where $\mu^2$ and $\zeta$ are constants. We also focus on this case and obtain the picture of accretion. The Higgs potential is the most widely used in particle physics. 
Now we redefine constants $\mu^2$ and $\zeta$ into a dimensionless form as \cite{Danila2019}
\begin{equation}\label{valpha}
v(\phi)=\alpha \phi^2+\beta \phi^4,
\end{equation}
where
\begin{equation}
\alpha=-\frac{1}{4}\biggl(\frac{ 2GnM_{BH} }{ c^2}\biggr)^2\mu^2,\ \ \ \ \ \ \beta=\frac{1}{2} \biggl(\frac{ 2GnM_{BH} }{ c^2}\biggr)^2\zeta^2.
\end{equation}
The Higgs-type potential yields four-parameter $(\alpha, \beta,\phi_0, u_0)$ solutions of the static gravitational field equations in HMPG. Authors of the article \cite{Danila2019} restrict their analysis investigating the role of constants $\alpha$ and $\beta$, while keeping  $\phi_0$ and $u_0$ fixed, and varying numerical values of $\alpha$ and $\beta$. However, we consider a wider range of cases:
\begin{enumerate}
\item We fix $u_0 = 10^{-8}, \phi_0 = 1, \beta = 10^{-10},$ and take the range of $\alpha =[-10^{-6}; -4\times10^{-5}]$. 
\item We fix $u_0 = 10^{-8}, \phi_0 = 1, \alpha = -10^{-10},$ and vary $\beta =[2\times 10^{-10}; 14\times10^{-10}]$.  
\item We study the case based on the Solar system data. We assume that the scalar field mass is defined as \cite{Capozziello2015, Dyadina2019}
\begin{equation}\label{ppn}
m_\varphi^2=[2V_0-V_\phi-(1+\phi)\phi V_{\phi \phi}]/3|_{\phi=\phi_0},
\end{equation} 
where subscript $\phi$ denotes the derivative with respect to the scalar field. Subsequently, $m_\varphi^2$ has a connection with $\alpha$ and $\beta$ parameters: 
\begin{equation}\label{hyggs}
m_\varphi^2=[- 4/3\alpha\phi_0 - 16/3\beta\phi_0^3 -10/3\beta\phi_0^4 ] \times 2\bigg(\frac{c^2}{2GM_{BH}}\bigg)^2.
\end{equation}
We save the connection between $u_0$ and $\phi_0$, which is known from PPN analysis as derivative of $\varphi=\frac{-2GM\phi_0e^{-m_\phi r}}{3c^2r}$ with respect to $r$. As a result, we get
\begin{equation}\label{u0V}
u_0=-\frac{2GM\phi_0e^{-m_\phi r}m_\phi}{3c^2r} - \frac{2GM\phi_0e^{-m_\phi r}}{3c^2r^2}. 
\end{equation}
In this case, we vary $\alpha =[-10^{-6}; -4\times10^{-5}]$ and fix $\beta = 10^{-20}$, $\phi_0=4\times10^{-5}$. 
\item We assume  $m_\varphi^2=-\mu^2$ by analogy with quantum field theoretical models. We choose the following set of parameters: $\alpha = [-10^{-6}; -4\times10^{-5}], \beta= 10^{-11}, \phi_0 = 4\times10^{-5}$. 
\item We also consider the case of large values of $\phi_0=[0.1;4]$ and fix $\alpha = -10^{-10}, \beta = 10^{-11}$, $m_\phi^2 = -\mu^2$.
\end{enumerate}

The first two cases are inspired by the work \cite{Danila2019}. However, we slightly extend the range of $\alpha$ and $\beta$ parameters to better illustrate  changes in accretion characteristics. The third case is based on  post-Newtonian analysis, relationships and constraints obtained from the solar system. The fourth and fifth instances include the connection $m_\varphi^2=-\mu^2$ that emerges from quantum field theoretical models, where $-\mu^2$ signifies the scalar field particle mass. By extension, we give the same meaning to this quantity in HMPG. Data from accelerator experiments suggest that the Higgs self-coupling constant $\zeta \approx 1/8$ for strong interactions. Nevertheless, the characteristics of the self-action of a scalar field in HMPG can differ markedly from those of Higgs bosons.

Further each of these cases is used to determine the characteristics of the accretion disk.

\subsubsection{Energy flux}
\begin{figure*}
		\begin{minipage}[h]{.45\textwidth}
			\includegraphics[width=\columnwidth]{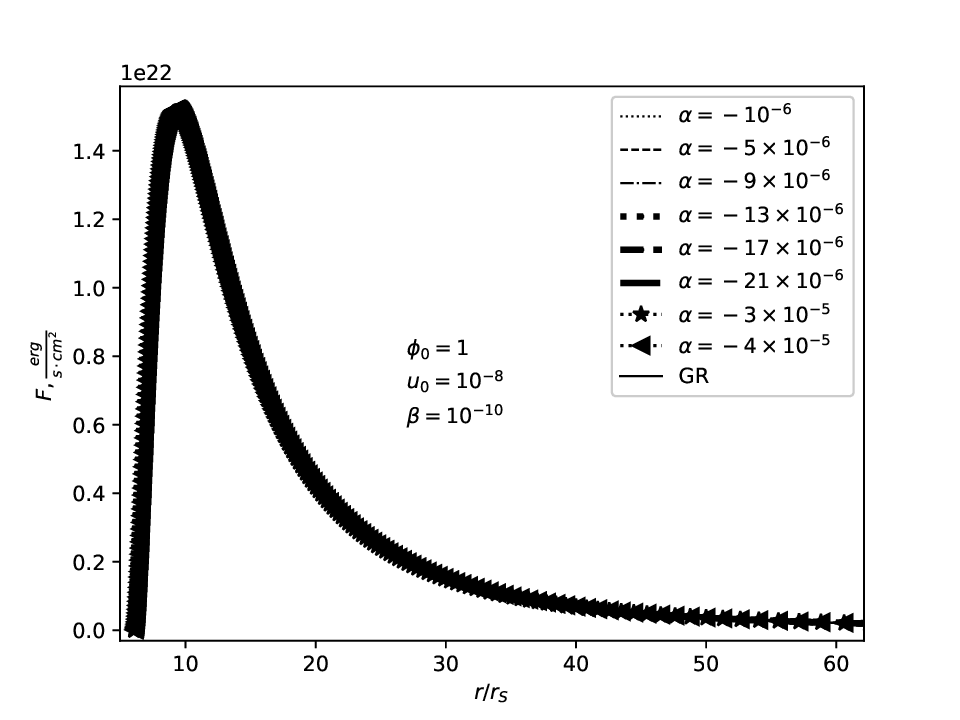} \\ a)
					\end{minipage}
\hfill
		\begin{minipage}[h]{.45\textwidth}
			{\includegraphics[width=\columnwidth]{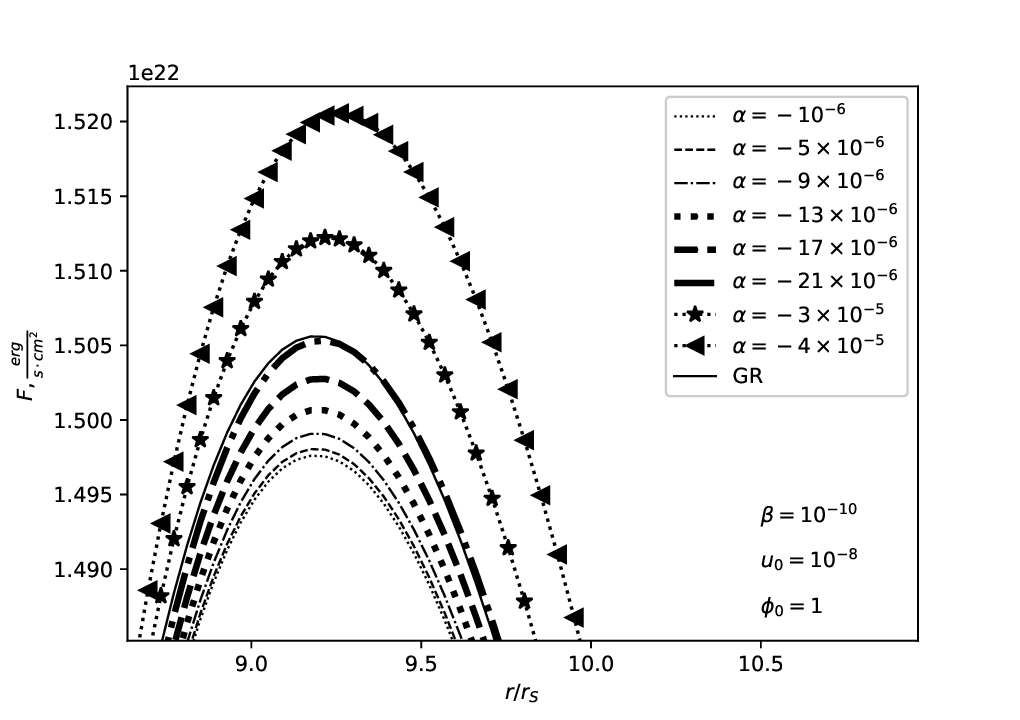}} \\ b)
				
		\end{minipage}

	\hfill
		\begin{minipage}[h]{.45\textwidth}
			\includegraphics[width=\columnwidth]{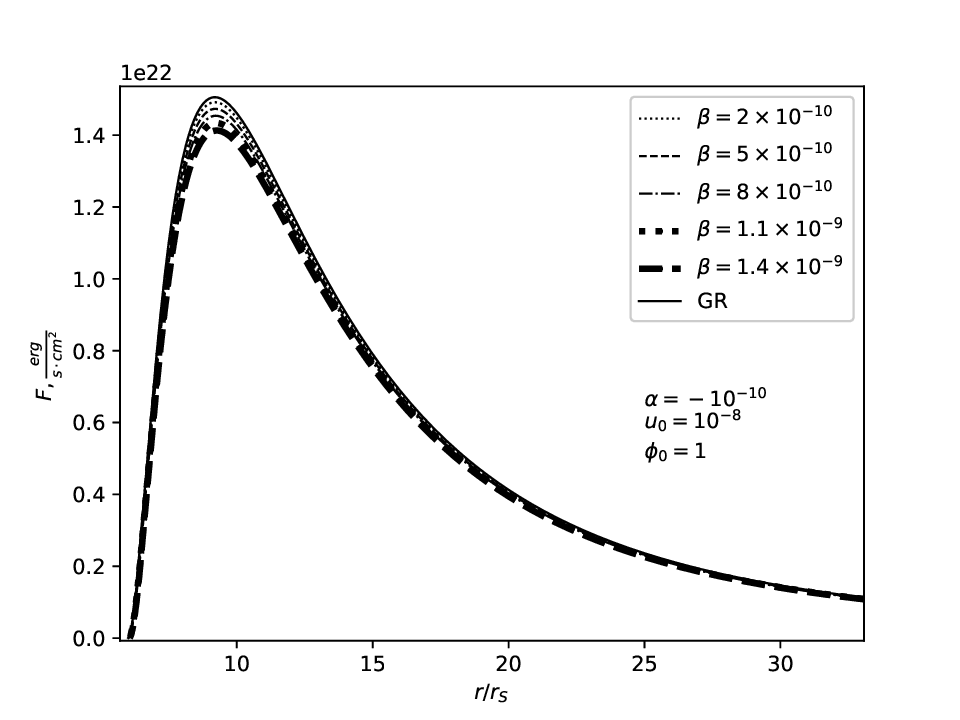} \\ c)
					\end{minipage}
\hfill
		\begin{minipage}[h]{.45\textwidth}
			{\includegraphics[width=\columnwidth]{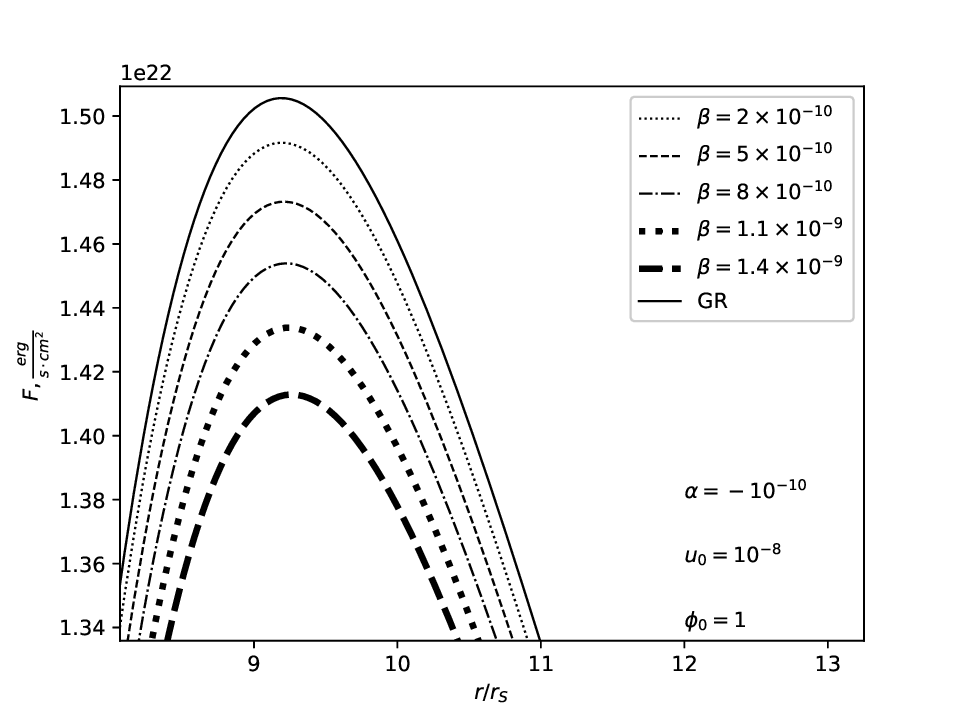}} \\ d)
		
		\end{minipage}
\caption{Hyggs-type potential case. The energy flux $F(r)$ of a disk around a static black hole with $\dot M = 2.21\times10^{18} g/s$ and $M=8.48 M_\odot$   as function of the normalized radial coordinate $r/r_s$.   	b) is zoom version a);	d) is zoom version c).}
\label{fig:fluxV1}	
		\end{figure*}

\begin{figure*}
		\begin{minipage}[h]{.45\textwidth}
			\includegraphics[width=\columnwidth]{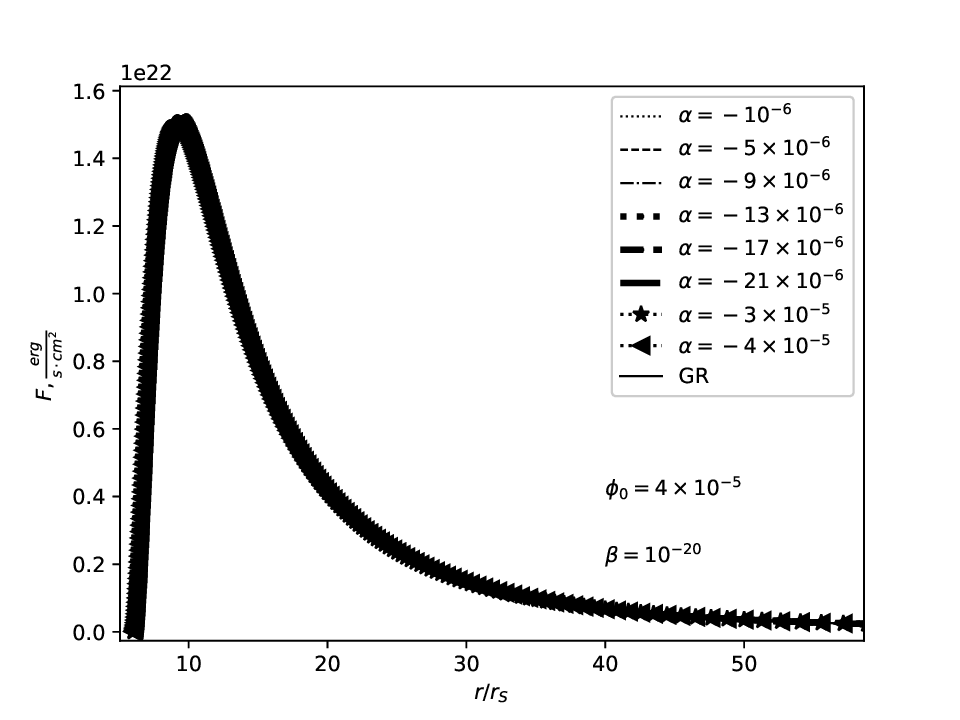} \\ a)
					\end{minipage}
\hfill
		\begin{minipage}[h]{.45\textwidth}
			{\includegraphics[width=\columnwidth]{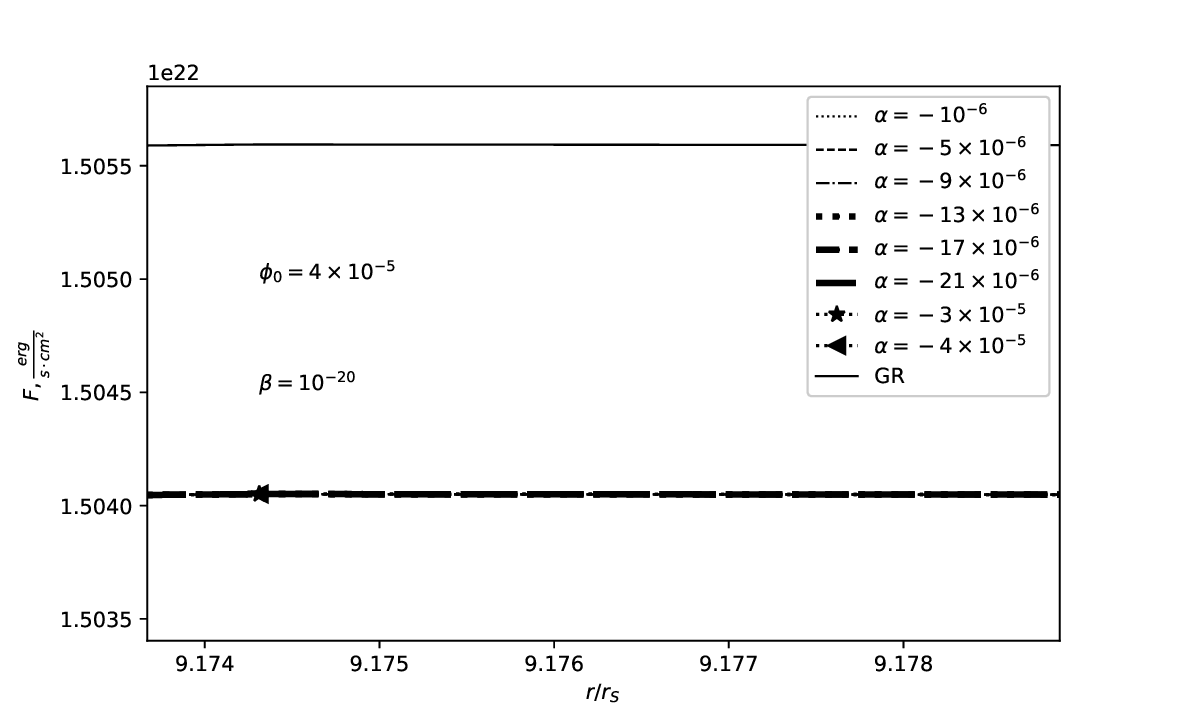}} \\ b)
	
		\end{minipage}
\hfill 
\begin{center}
\begin{minipage}[h]{.45\textwidth}
			{\includegraphics[width=\columnwidth]{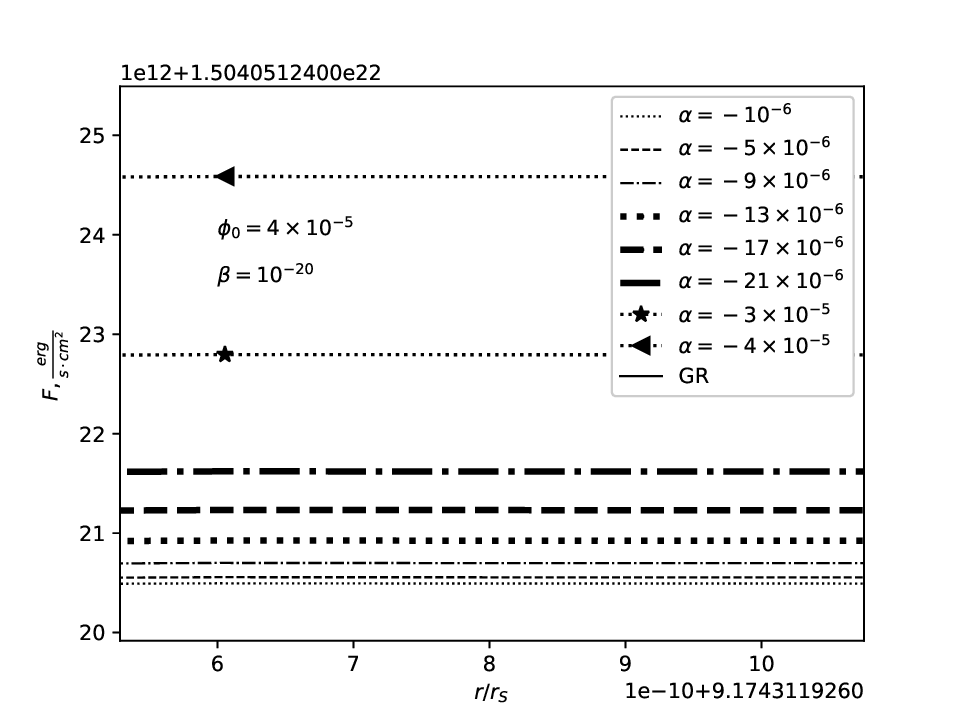}} \\ c)
			
		\end{minipage}
\end{center}
\caption{Hyggs-type potential case. The energy flux $F(r)$ of a disk around a static black hole with $\dot M = 2.21\times10^{18} g/s$ and $M=8.48 M_\odot$   as function of the normalized radial coordinate $r/r_s$.   	The connections  (\ref{hyggs}) and (\ref{u0V}) are taken into accaunt. b) and c) are zoom versions a).}

\label{fig:fluxV3}
\end{figure*}

\begin{figure*}
	
		\begin{minipage}[h]{.45\textwidth}
			\includegraphics[width=\columnwidth]{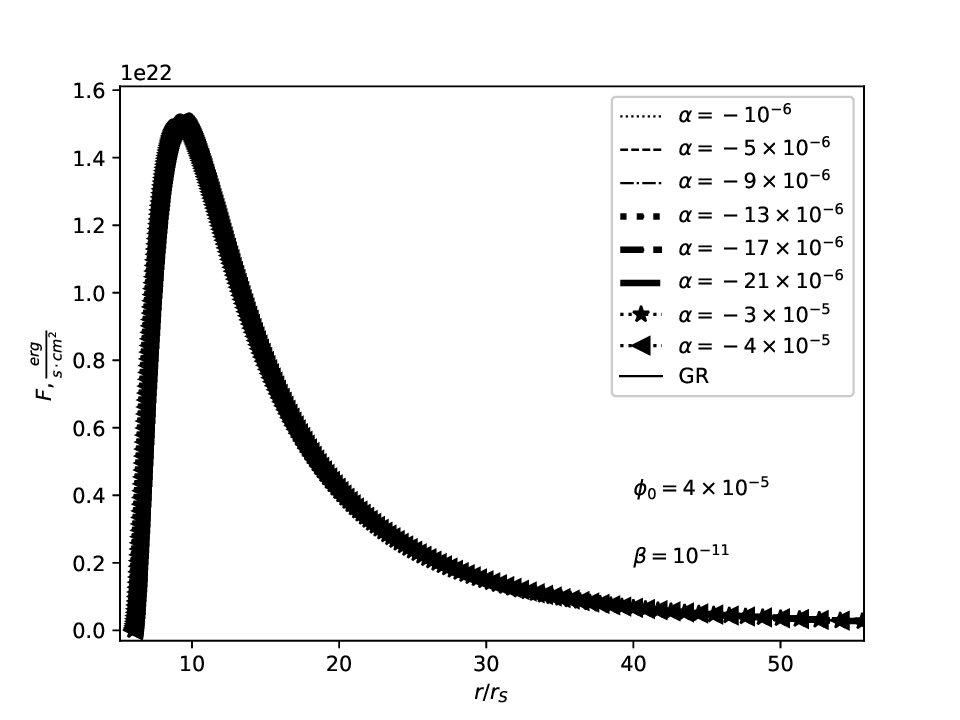} \\ a)
					\end{minipage}
\hfill
		\begin{minipage}[h]{.45\textwidth}
			{\includegraphics[width=\columnwidth]{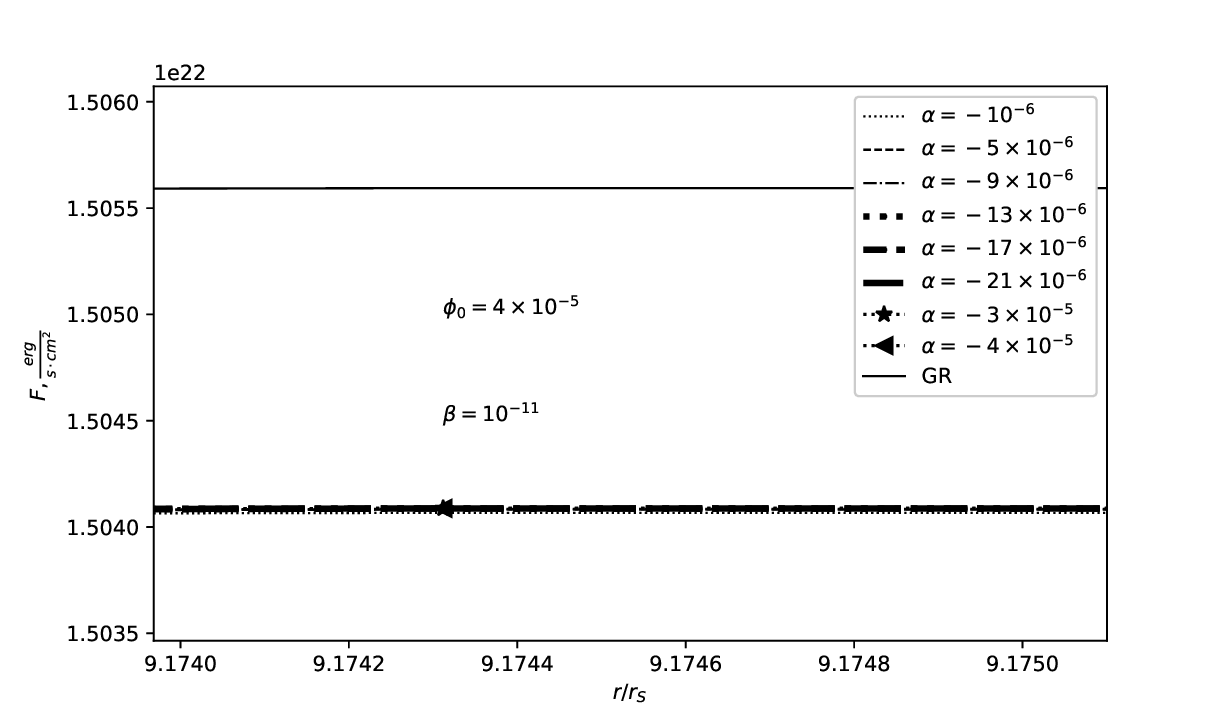}} \\ b)
		
		\end{minipage}
\hfill 
\begin{minipage}[h]{.45\textwidth}
			{\includegraphics[width=\columnwidth]{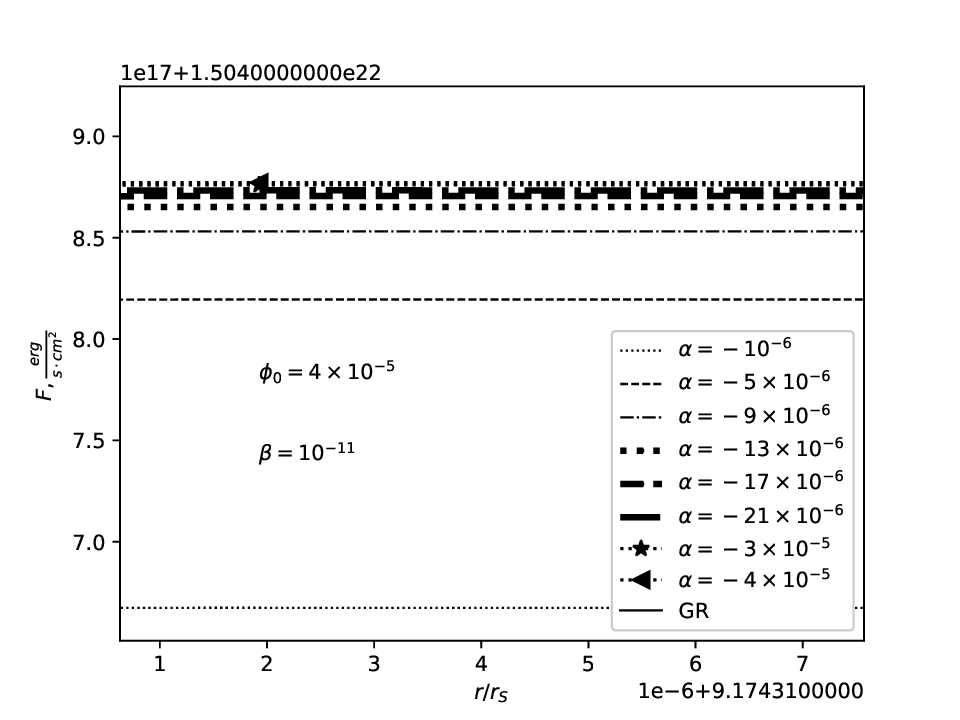}} \\ c)
		
		\end{minipage}
\hfill 
\begin{minipage}[h]{.45\textwidth}
			{\includegraphics[width=\columnwidth]{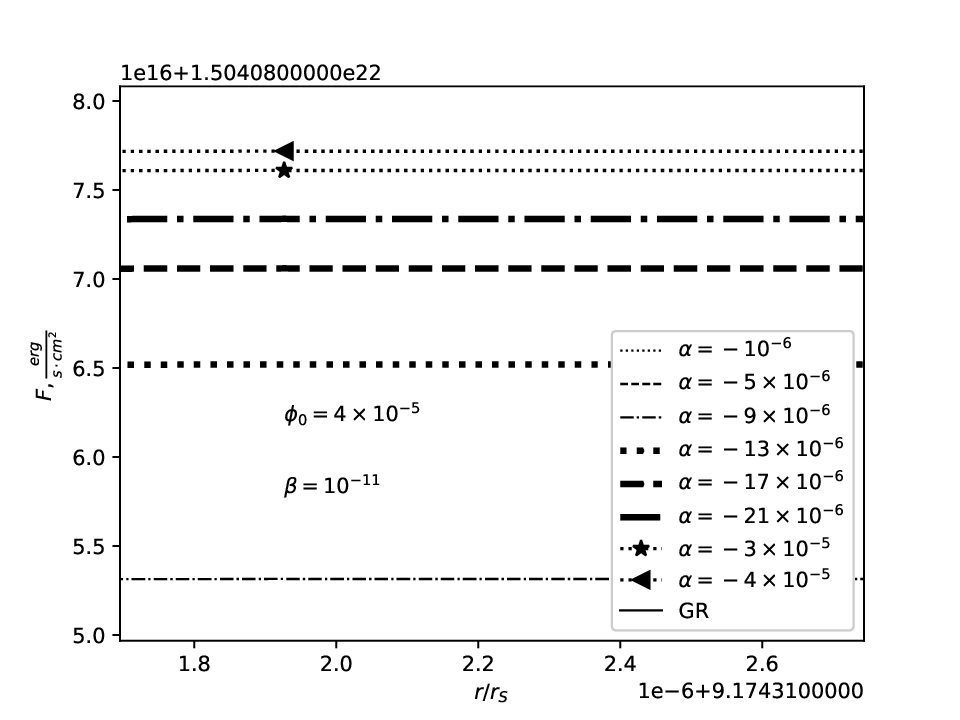}} \\ d)
	
		\end{minipage}

\hfill
\begin{center}
\begin{minipage}[h]{.55\textwidth}
			{\includegraphics[width=\columnwidth]{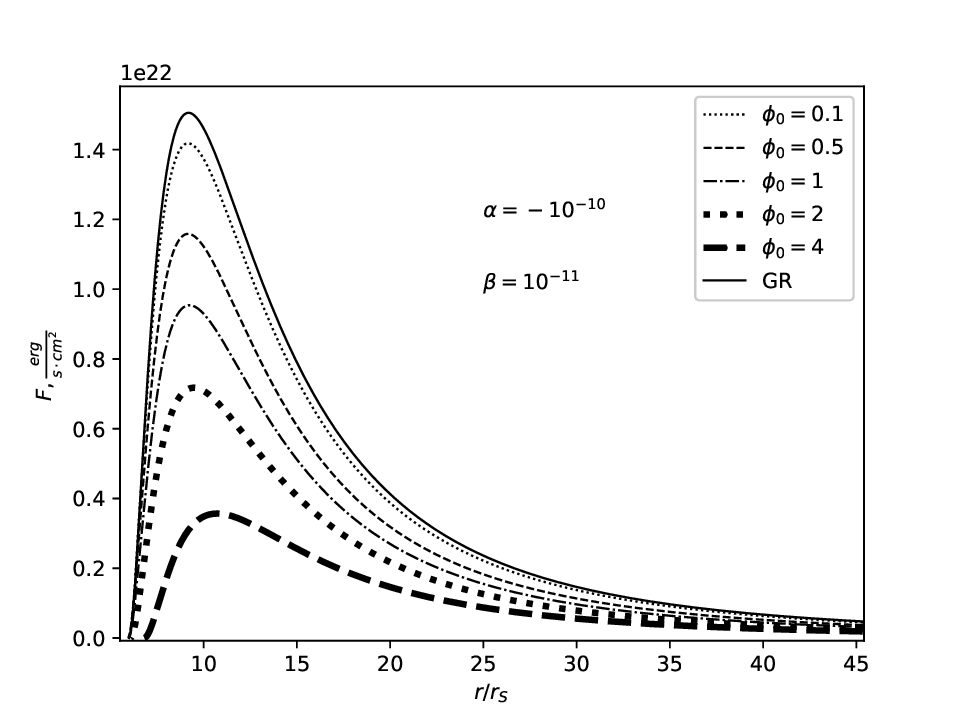}} \\ e)

	\end{minipage}
	\end{center}
		\caption{Hyggs-type potential case. The energy flux $F(r)$ of a disk around a static black hole with $\dot M = 2.21\times10^{18} g/s$ and $M=8.48 M_\odot$   as function of the normalized radial coordinate $r/r_s$.   The connection  $m_\varphi^2=-\mu^2$ is taken into accaunt.	b), c), d) are zoom versions of a).}
		\label{fig:fluxV4}
	\end{figure*}

\begin{enumerate}
\item This case is illustrated in Fig. \ref{fig:fluxV1}a, \ref{fig:fluxV1}b. We can observe that as the modulus $\alpha$ increases, the energy flux maximum also increases. Notably,  at $\alpha=-10^{-5}$, it exceeds the Schwarzschild result.  The maximum deviation for curves both above and below the  Schwarzschild one is 0.95\% and 0.47\%, respectively.

\item When we vary the $\beta$ parameter, no curve exceeds the Schwarzschild result. Furthermore, as $\beta$ increases, the maximum of the energy flux decreases. This result is displayed in  Fig. \ref{fig:fluxV1}c, \ref{fig:fluxV1}d.

\item As shown in Fig. \ref{fig:fluxV3} the energy flux maximum increases with the growth of the modulus $\alpha$. However, the change in the $\beta$ parameter, while  $\alpha$  remaines fixed, does not affect the position of the curve. This is due to the fact that $\beta$ is multiplied by greater powers of $\phi_0$ than $\alpha$ (see eq. \ref{valpha}), which means that its contribution is suppressed. 

\item As in previous cases, with increasing modulus $\alpha$, an increase in the energy flux maximum is observed (see Fig. \ref{fig:fluxV4}a, \ref{fig:fluxV4}b, \ref{fig:fluxV4}c, \ref{fig:fluxV4}d). However, in none of the cases is there an excess of the Schwarzschild curve. The maximum deviation from the Schwarzschild curve in this case is 0.01\%. Similar to the previous case, changing $\beta$ with a fixed $\alpha$ does not contribute to the shift of the energy flux curve.

\item As $\phi_0$ increases, the maximum value of the energy flux decreases. The maximum deviation is 80\% from the GR curve. This case is illustrated in Fig. \ref{fig:fluxV4}e.
\end{enumerate}
	


\subsubsection{Emission spectra.}
\begin{figure*}
		\begin{minipage}[h]{.45\textwidth}
			\includegraphics[width=\columnwidth]{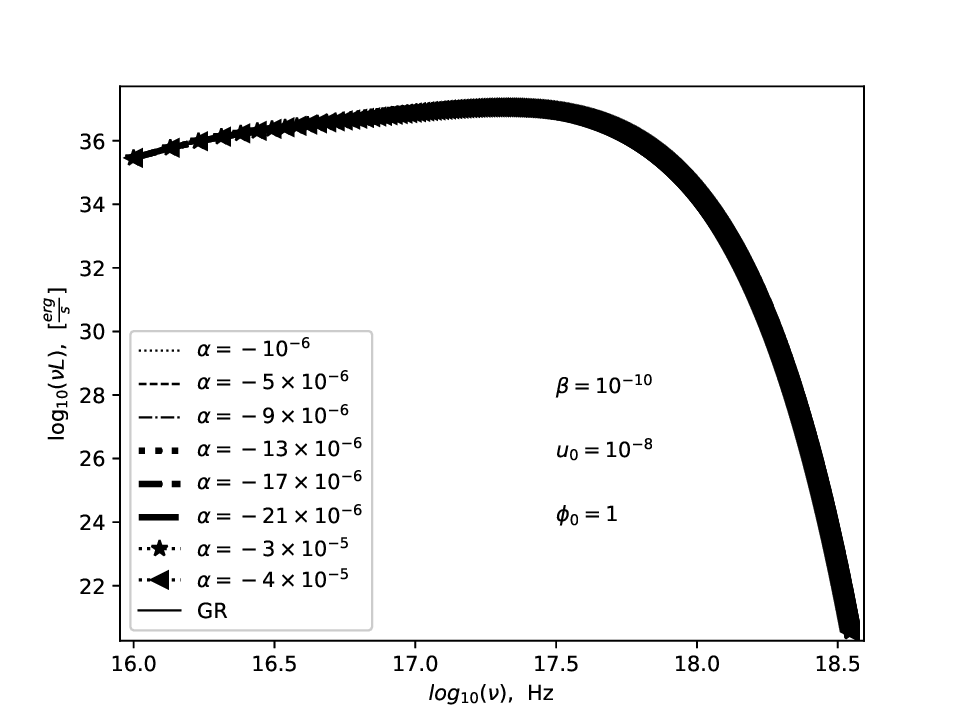} \\ a)
					\end{minipage}
\hfill
		\begin{minipage}[h]{.45\textwidth}
			{\includegraphics[width=\columnwidth]{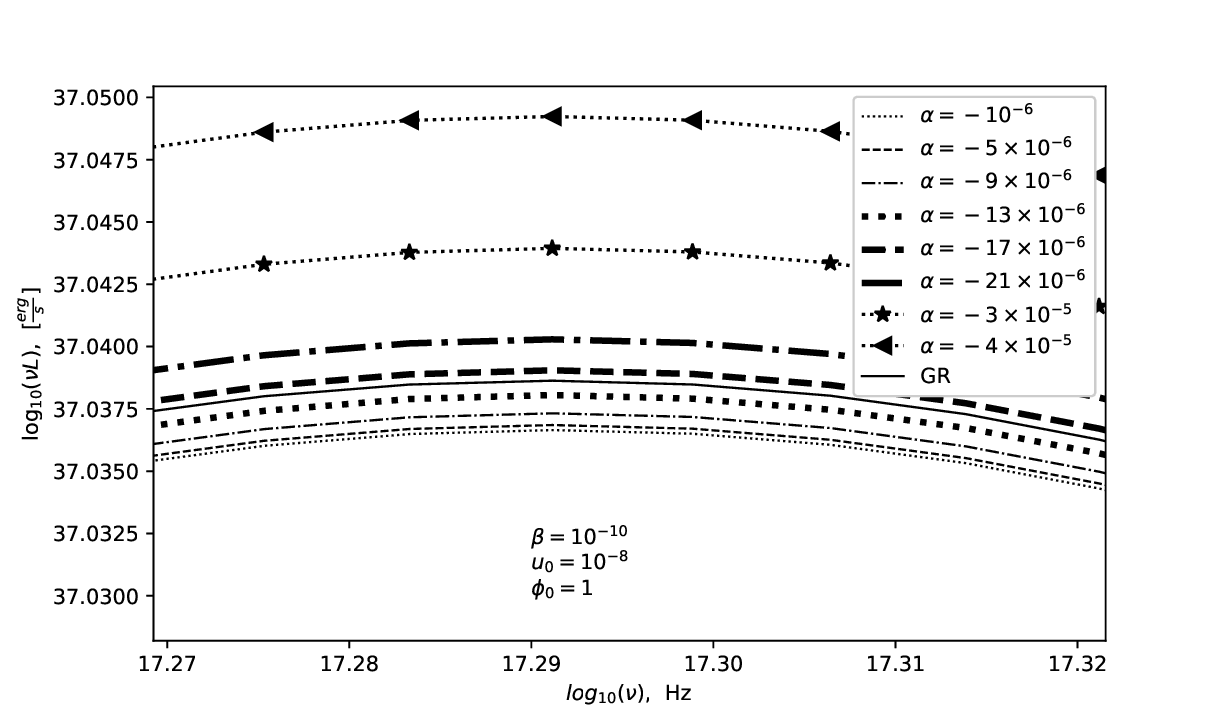}} \\ b)
		\end{minipage}
\hfill
		\begin{minipage}[h]{.45\textwidth}
			\includegraphics[width=\columnwidth]{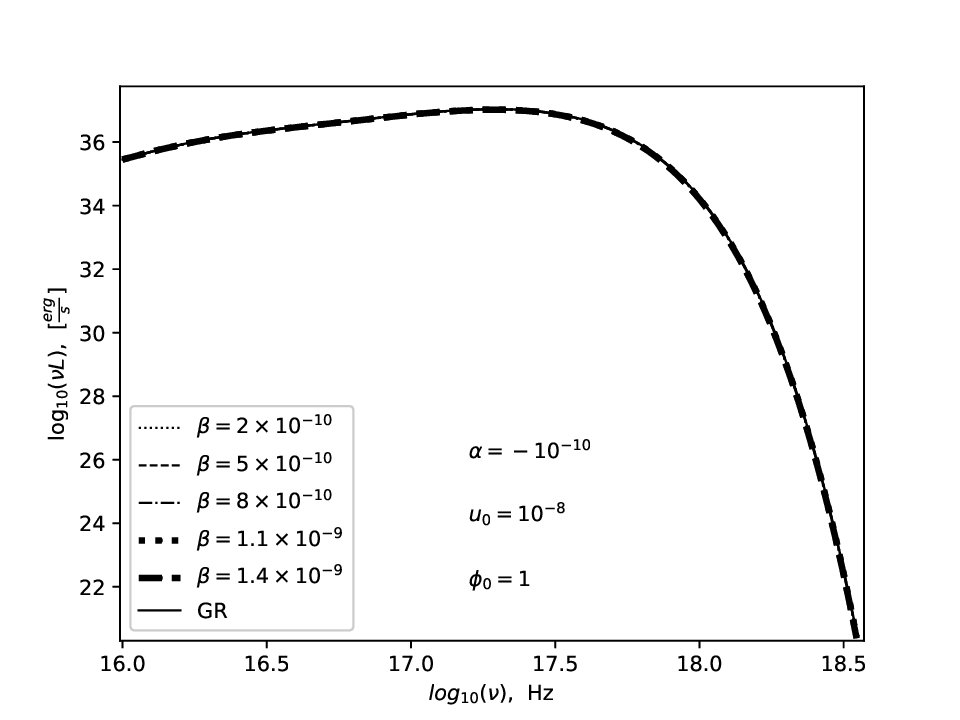} \\ c)
					\end{minipage}
\hfill
		\begin{minipage}[h]{.45\textwidth}
			{\includegraphics[width=\columnwidth]{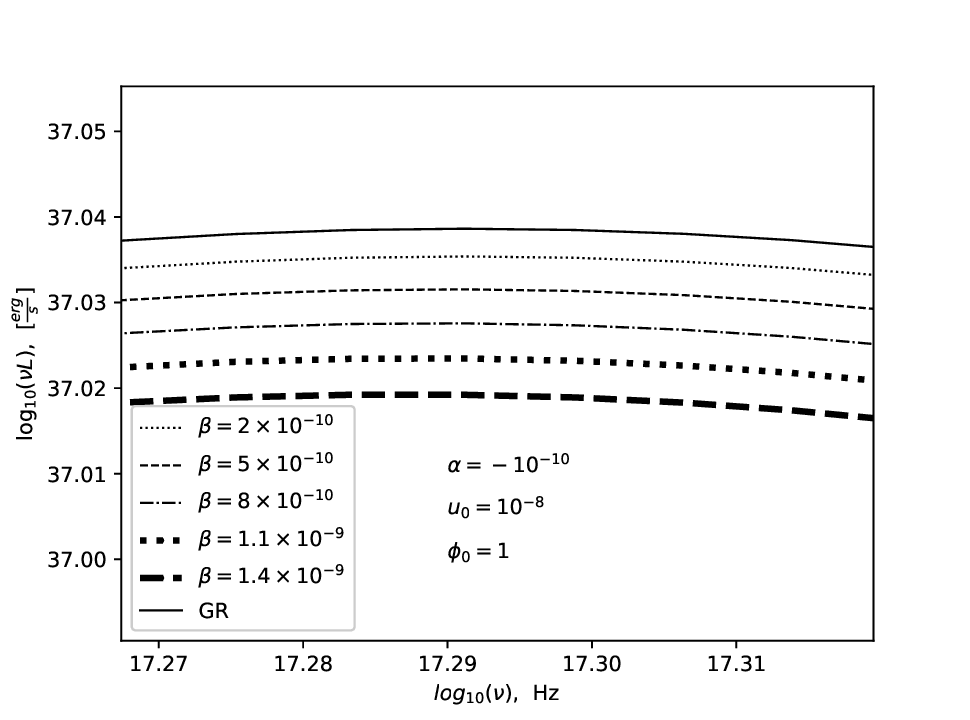}} \\ d)
		
		\end{minipage}
\caption{Hyggs-type potential case. The emission spectrum $\nu L(\nu)$ of the accretion disk around a static black hole with $\dot M = 2.21\times10^{18} g/s$ and $M=8.48 M_\odot$   as function  of frequency $\nu$. b) is zoom version of a); d) is zoomversion of c).				}
		\label{fig:LumV1}	
	\end{figure*}

\begin{figure*}
		\begin{minipage}[h]{.45\textwidth}
			\includegraphics[width=\columnwidth]{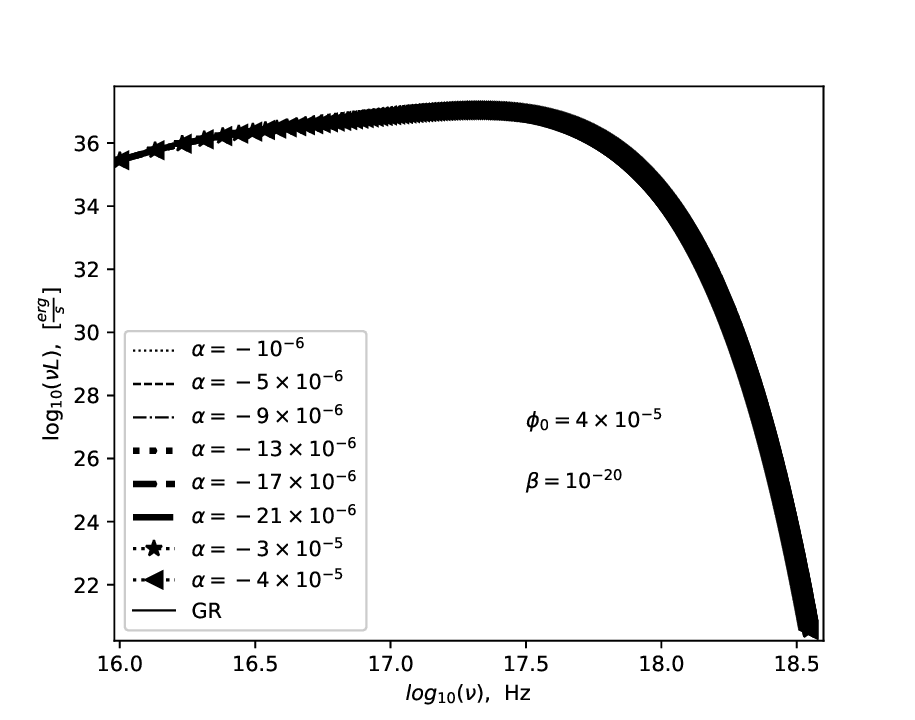} \\ a)
					\end{minipage}
\hfill
		\begin{minipage}[h]{.45\textwidth}
			{\includegraphics[width=\columnwidth]{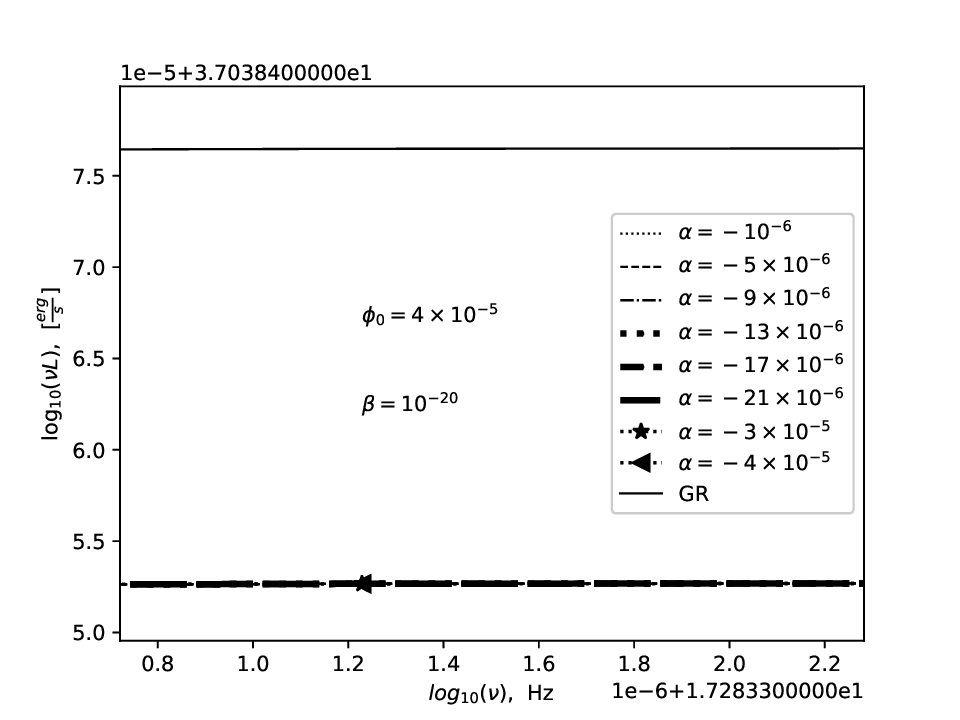}} \\ b)
			
		\end{minipage}
\hfill 
\begin{center}
\begin{minipage}[h]{.45\textwidth}
			{\includegraphics[width=\columnwidth]{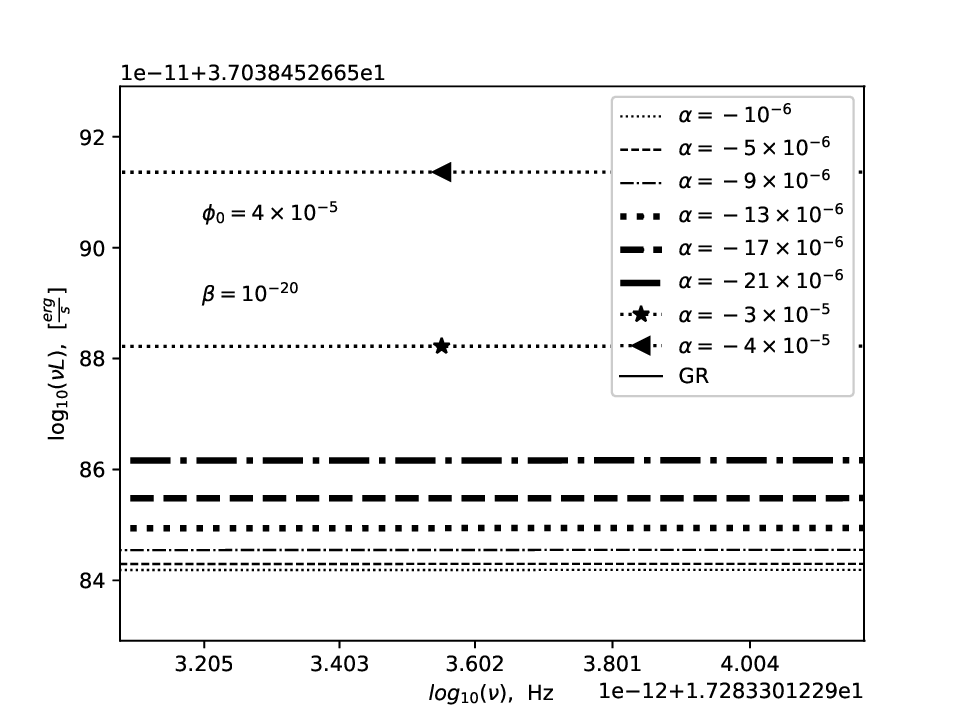}} \\ c)
			
		\end{minipage}
\end{center}
		\caption{Hyggs-type potential case. The emission spectrum $\nu L(\nu)$ of the accretion disk around a static black hole with $\dot M = 2.21\times10^{18} g/s$ and $M=8.48 M_\odot$  as function  of frequency $\nu$. The connections (\ref{hyggs}) and (\ref{u0V})  are taken into accaunt. b) and c) are zoom versions of a). 	}
		
		\label{fig:LumV3}
	
	\end{figure*}

\begin{figure*}
		\begin{minipage}[h]{.45\textwidth}
			\includegraphics[width=\columnwidth]{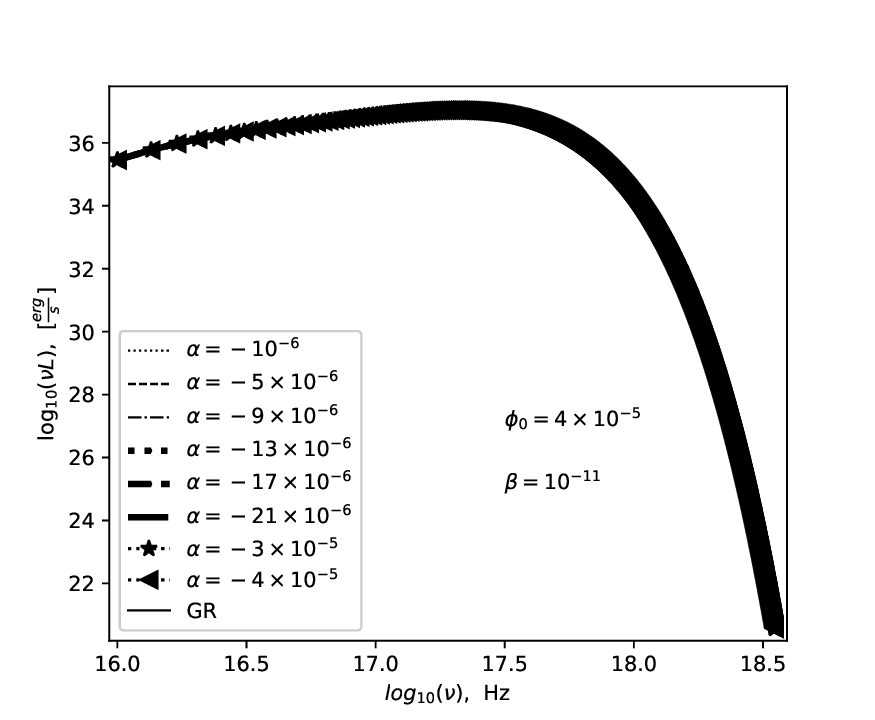} \\ a)
					\end{minipage}
\hfill
		\begin{minipage}[h]{.45\textwidth}
			{\includegraphics[width=\columnwidth]{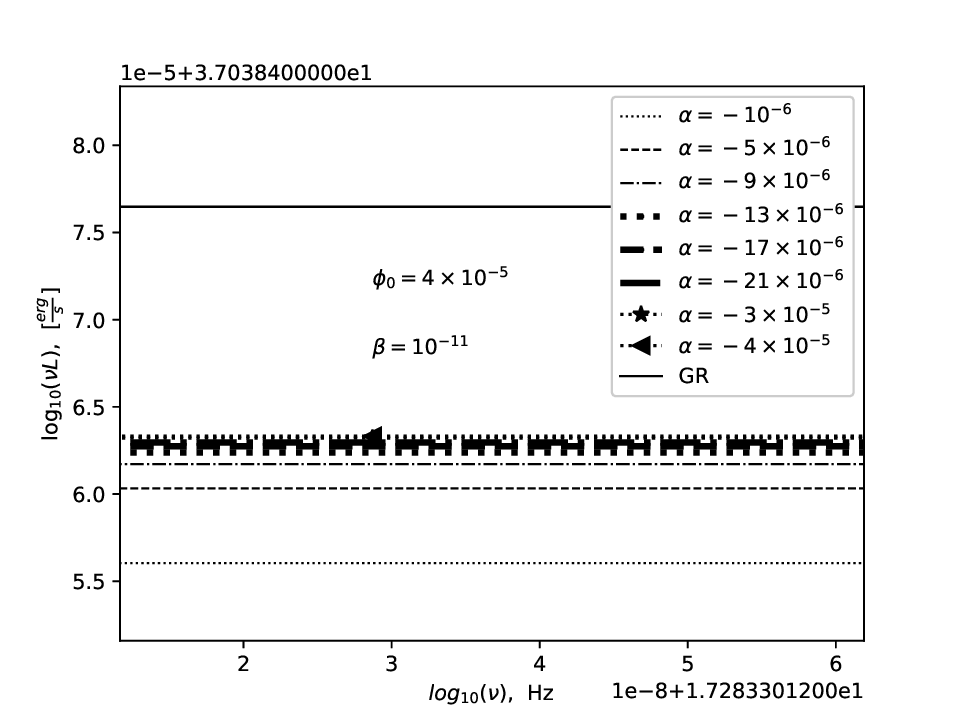}} \\ b)
			
		\end{minipage}

	\hfill
\begin{center}
		\begin{minipage}[h]{.45\textwidth}
			{\includegraphics[width=\columnwidth]{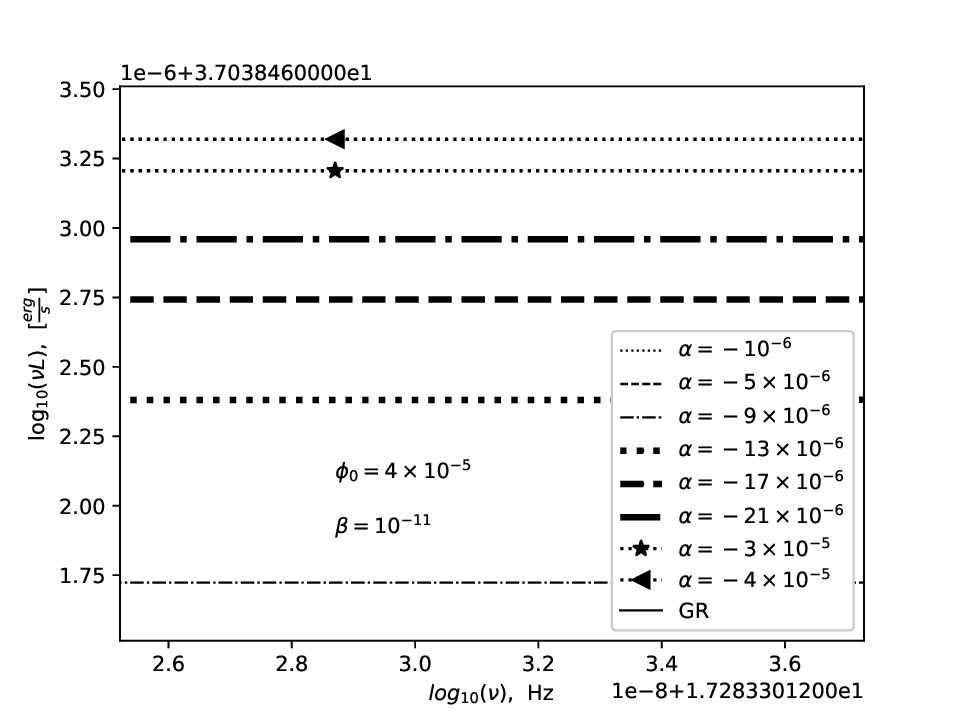}} \\ c)
			
		\end{minipage}
\end{center}	
\hfill

\begin{minipage}[h]{.45\textwidth}
			{\includegraphics[width=\columnwidth]{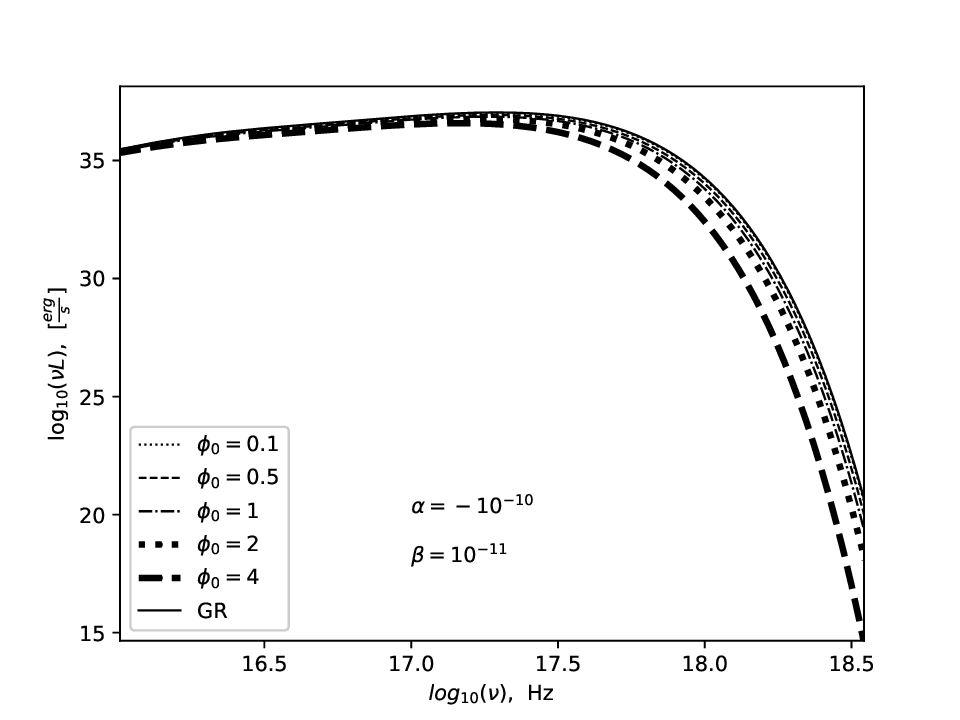}} \\ d)

	\end{minipage}
\hfill
\begin{minipage}[h]{.45\textwidth}
			{\includegraphics[width=\columnwidth]{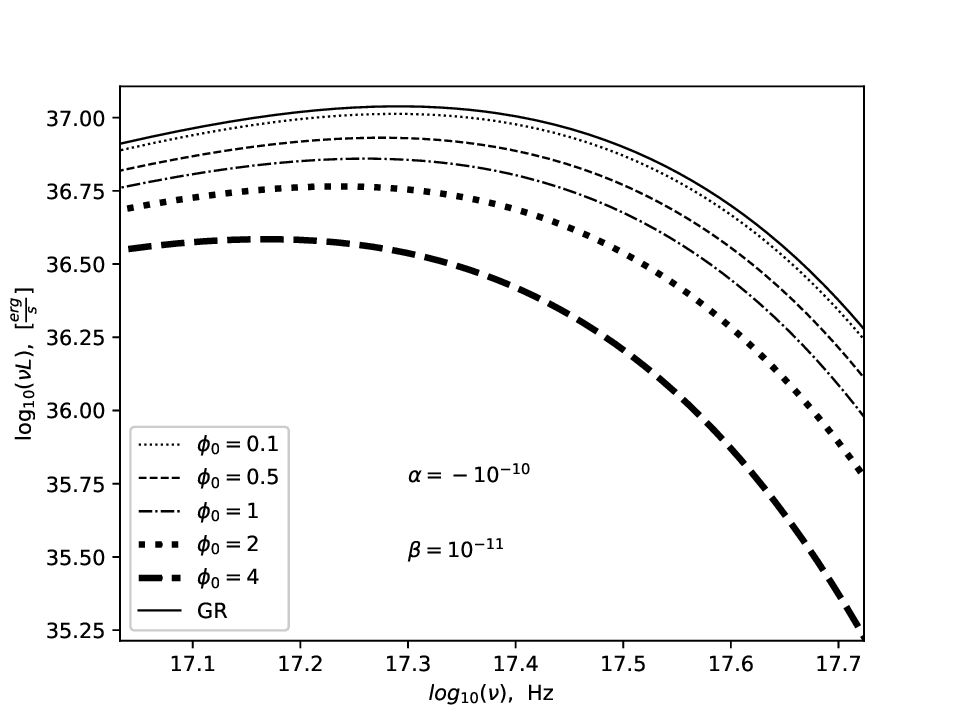}} \\ e)
	\end{minipage}

			\caption{Hyggs-type potential case. The emission spectrum $\nu L(\nu)$ of the accretion disk around a static black hole with $\dot M = 2.21\times10^{18} g/s$ and $M=8.48 M_\odot$  as function  of frequency $\nu$. The connection $m_\varphi^2=-\mu^2$ is taken into accaunt. b) and c) are zoom versions of a); e) is zoom version of d).		}
	\label{fig:LumV4}
	\end{figure*}

In this subsection we present the spectral energy distribution of the disk radiation around the black holes for the general relativistic case, and for the hybrid metric-Palatini gravity.
\begin{enumerate}
\item In Fig. \ref{fig:LumV1}a, \ref{fig:LumV1}b we vary $\alpha$ while  fixing other parameters. The luminosity maximum exceeds Schwarzschild predictions at $\alpha=-1.5\times10^{-5}$. Thus, the luminosity changes faster than the energy flux when we vary the $\alpha$ parameter. 
\item With a fixed $\alpha$ parameter and varying $\beta$ parameter, we see the same pattern as for the energy flux: with increasing $\beta$, the luminosity decreases (see Fig. \ref{fig:LumV1}c, \ref{fig:LumV1}d). 
\item Taking into account the connection \eqref{hyggs} between the mass of the scalar field $m_\phi$ and the parameters $\alpha$ and $\beta$, the picture does not differ from the energy flux. As the modulus $\alpha$ grows, the luminosity maximum increases. Changes in the $\beta$ parameter with a fixed $\alpha$ do not affect the position of the curve. This case is illustrated in Fig. \ref{fig:LumV3}.
\item In Fig. \ref{fig:LumV4}a, \ref{fig:LumV4}b, \ref{fig:LumV4}c,  we can see that the connection between mass of the scalar field and the parameter of potential  $m_\varphi^2=-\mu^2$ yields an increase in luminosity as the modulus of $\alpha$ increases. Varying the $\beta$ parameter in this case also has no effect. 
\item As $\phi$ varies, the largest values of $\phi$ produce the greatest deviation from the Schwarzschild curve. However, in this case, the maximum deviation from GR is only 3\% (see Fig. \ref{fig:LumV4}d, \ref{fig:LumV4}e).
\end{enumerate}

\subsubsection{Temperature.}
\begin{figure*}
		\begin{minipage}[h]{.45\textwidth}
			\includegraphics[width=\columnwidth]{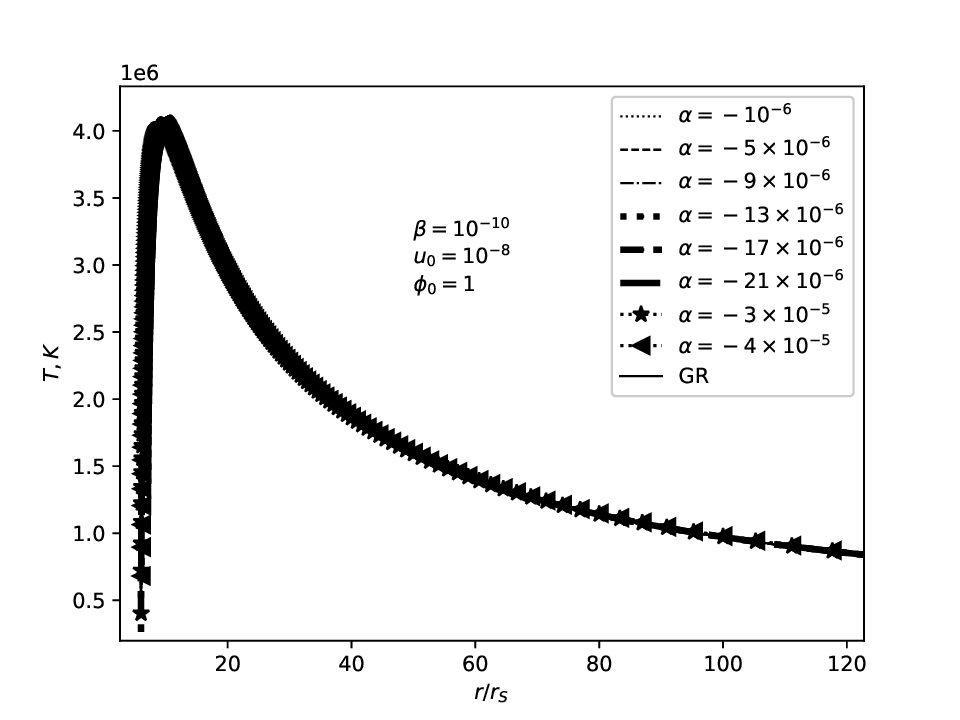} \\ a)
					\end{minipage}
\hfill
		\begin{minipage}[h]{.45\textwidth}
			{\includegraphics[width=\columnwidth]{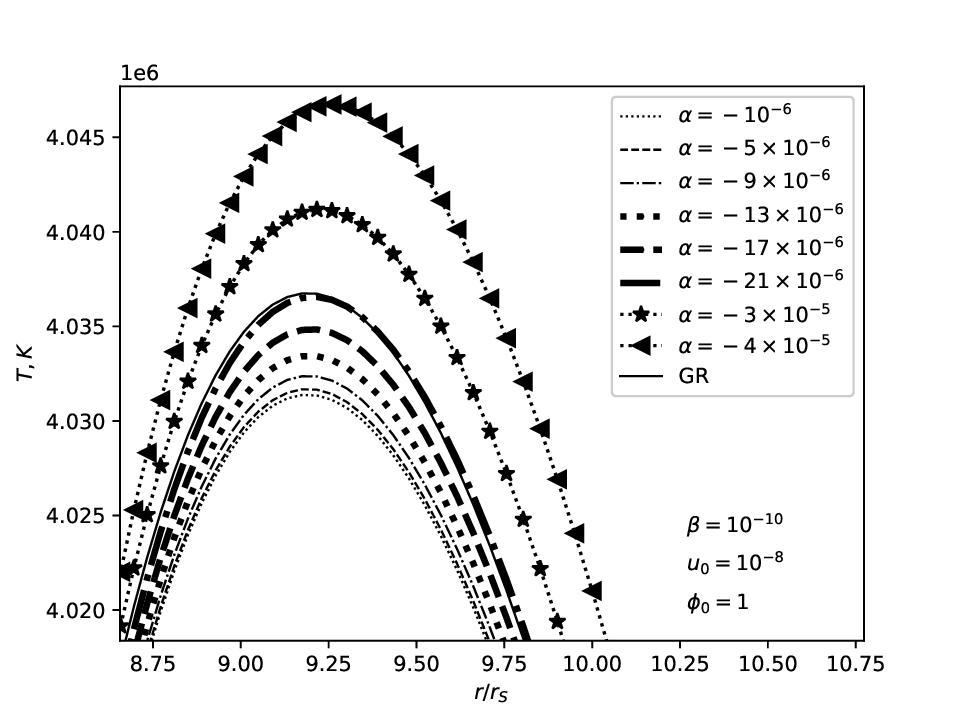}} \\ b)

		\end{minipage}
\hfill

		\begin{minipage}[h]{.45\textwidth}
			\includegraphics[width=\columnwidth]{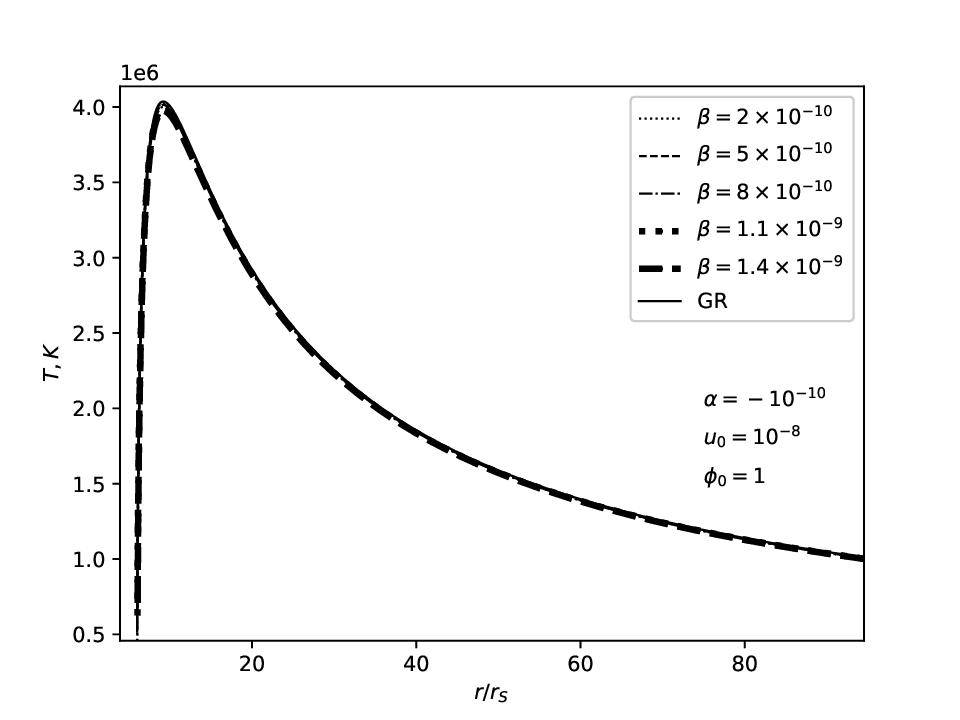} \\ c)
					\end{minipage}
\hfill
		\begin{minipage}[h]{.45\textwidth}
			{\includegraphics[width=\columnwidth]{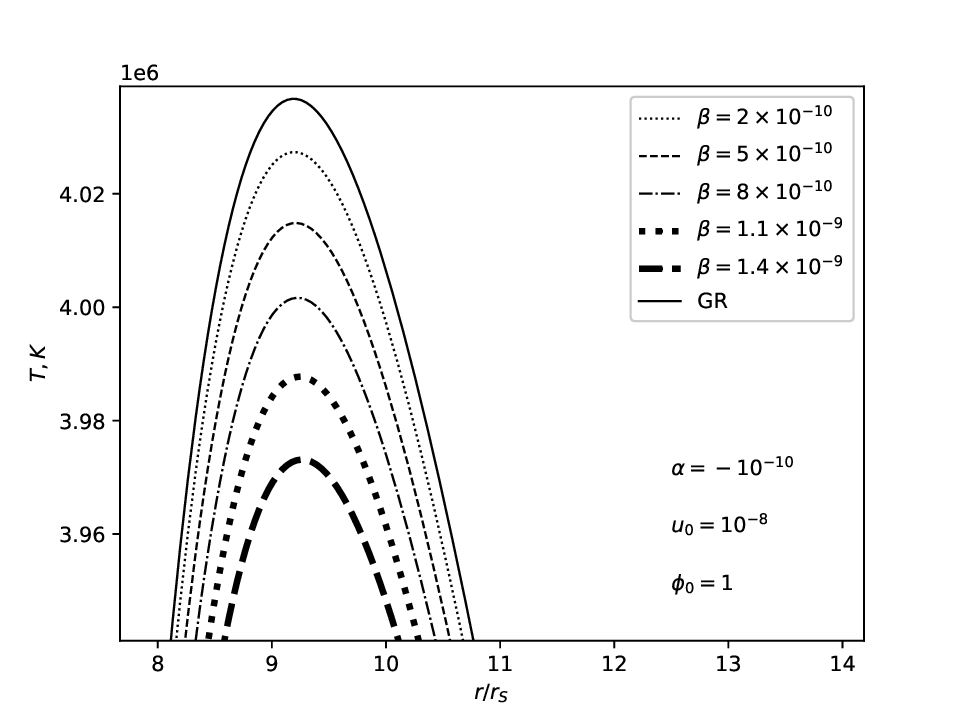}} \\ d)
			
		\end{minipage}
		\caption{Hyggs-type potential case.  The temperature distribution $T(r)$ of a disk around a static black hole with $\dot M = 2.21\times10^{18} g/s$ and $M=8.48 M_\odot$   as function of the normalized radial coordinate $r/r_s$. b) is zoom version of a), d) is zoom version of c). 		}
		\label{fig:TV1}
	
	\end{figure*}

\begin{figure*}
		\begin{minipage}[h]{.45\textwidth}
			\includegraphics[width=\columnwidth]{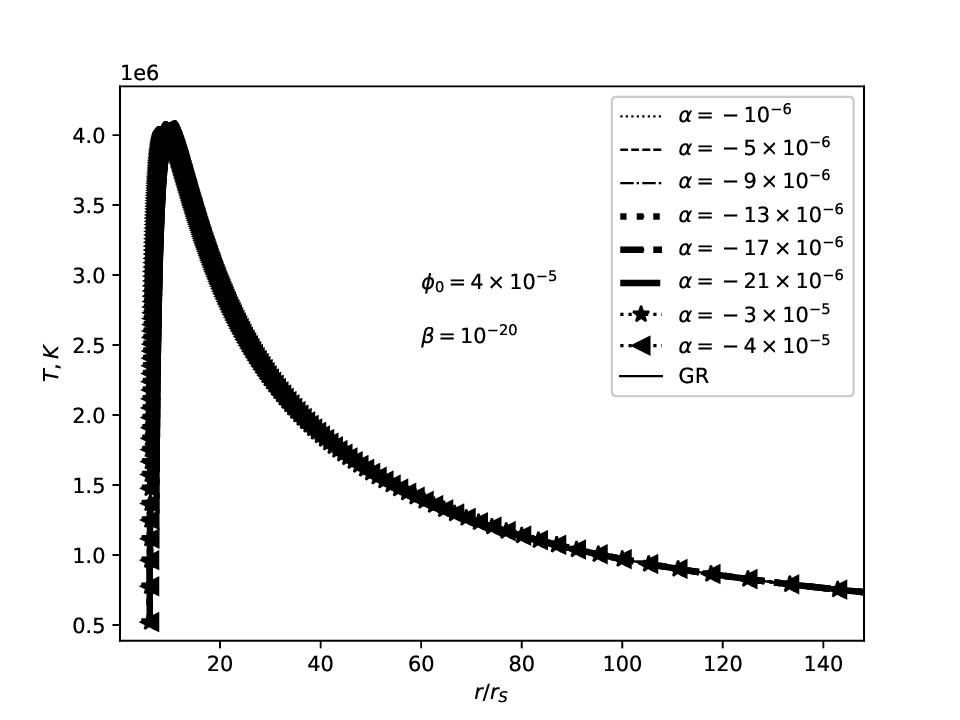} \\ a)
					\end{minipage}
\hfill
		\begin{minipage}[h]{.45\textwidth}
			{\includegraphics[width=\columnwidth]{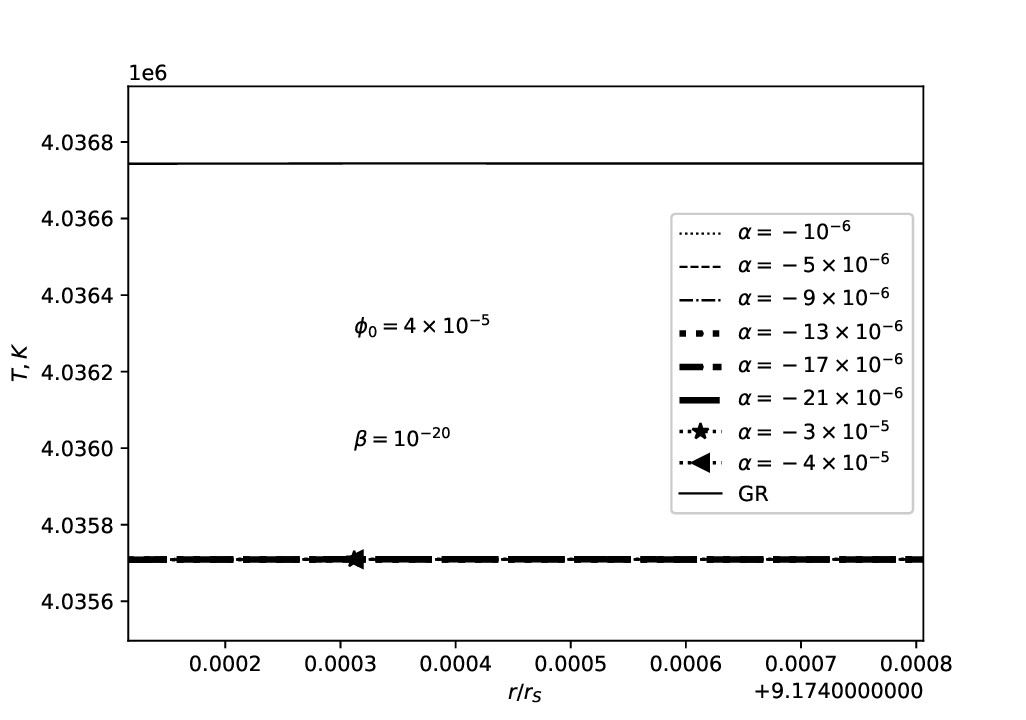}} \\ b)
			
		\end{minipage}
\hfill 
\begin{center}
\begin{minipage}[h]{.45\textwidth}
			{\includegraphics[width=\columnwidth]{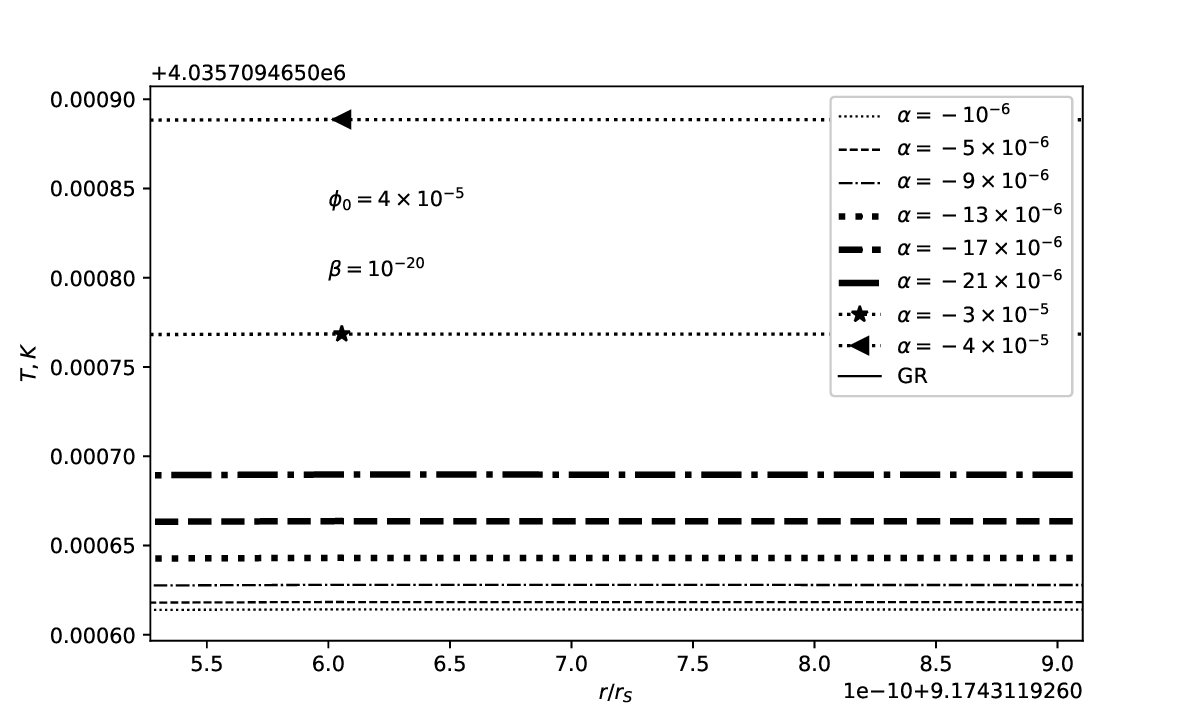}} \\ c)
			
		\end{minipage}
\end{center}
		\caption{Hyggs-type potential case.  The temperature distribution $T(r)$ of a disk around a static black hole with $\dot M = 2.21\times10^{18} g/s$ and $M=8.48 M_\odot$   as function of the normalized radial coordinate $r/r_s$. The connections (\ref{hyggs}) and (\ref{u0V}) are taken into accaunt. b) and c) are zoom versions of a).	}
		
		\label{fig:TV3}
	
	\end{figure*}

\begin{figure*}
		\begin{minipage}[h]{.45\textwidth}
			\includegraphics[width=\columnwidth]{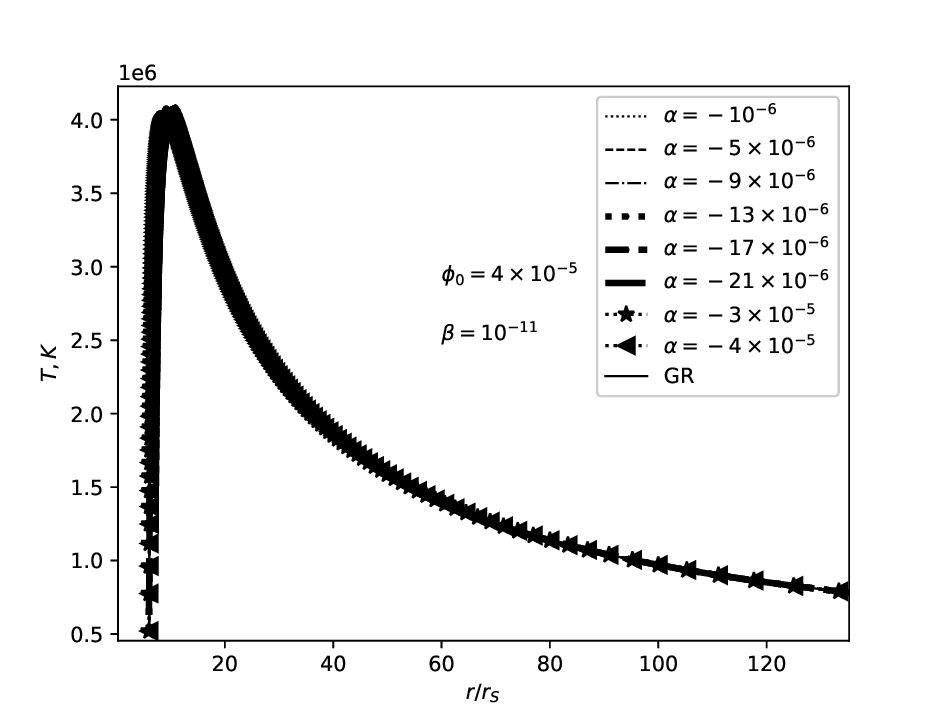} \\ a)
					\end{minipage}
\hfill
		\begin{minipage}[h]{.45\textwidth}
			{\includegraphics[width=\columnwidth]{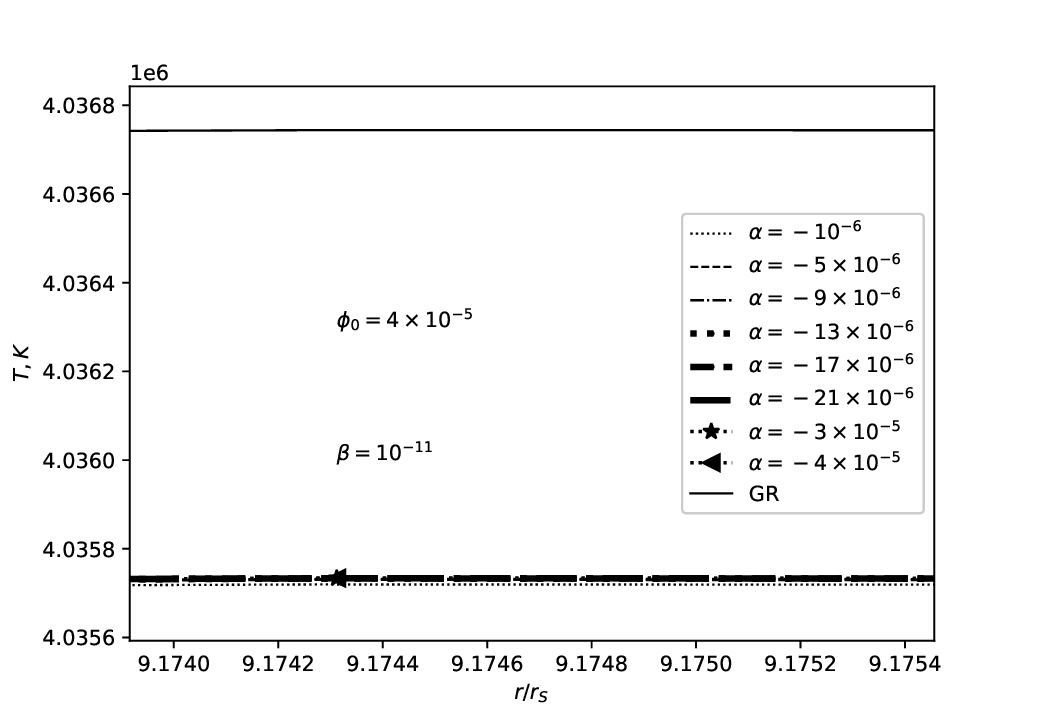}} \\ b)
			
		\end{minipage}
\hfill 
\begin{minipage}[h]{.45\textwidth}
			{\includegraphics[width=\columnwidth]{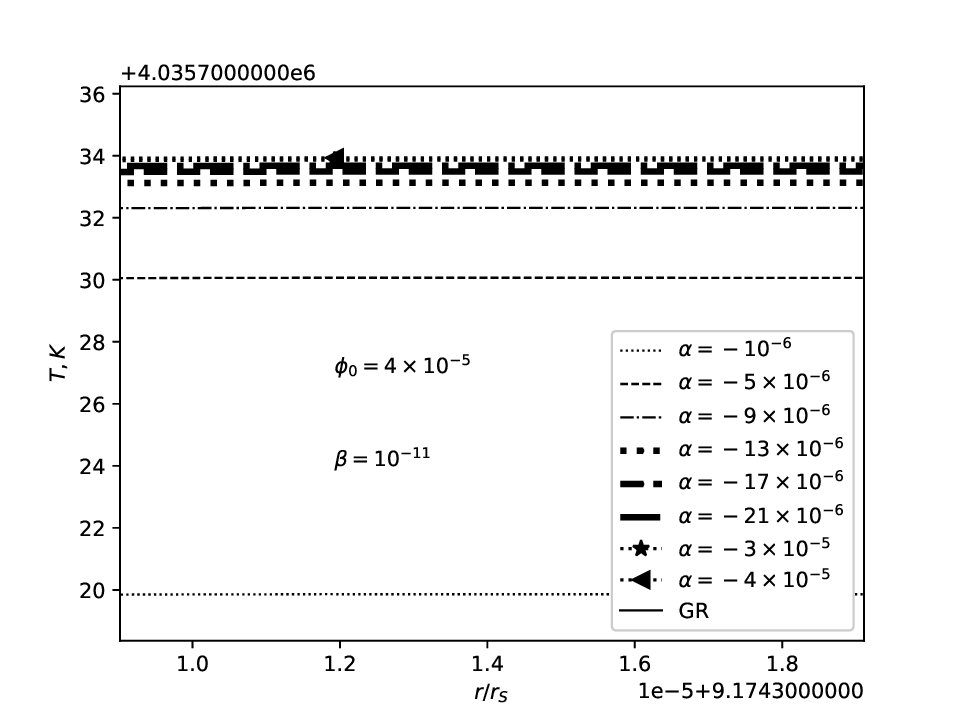}} \\ c)
			
		\end{minipage}
\hfill 
\begin{minipage}[h]{.45\textwidth}
			{\includegraphics[width=\columnwidth]{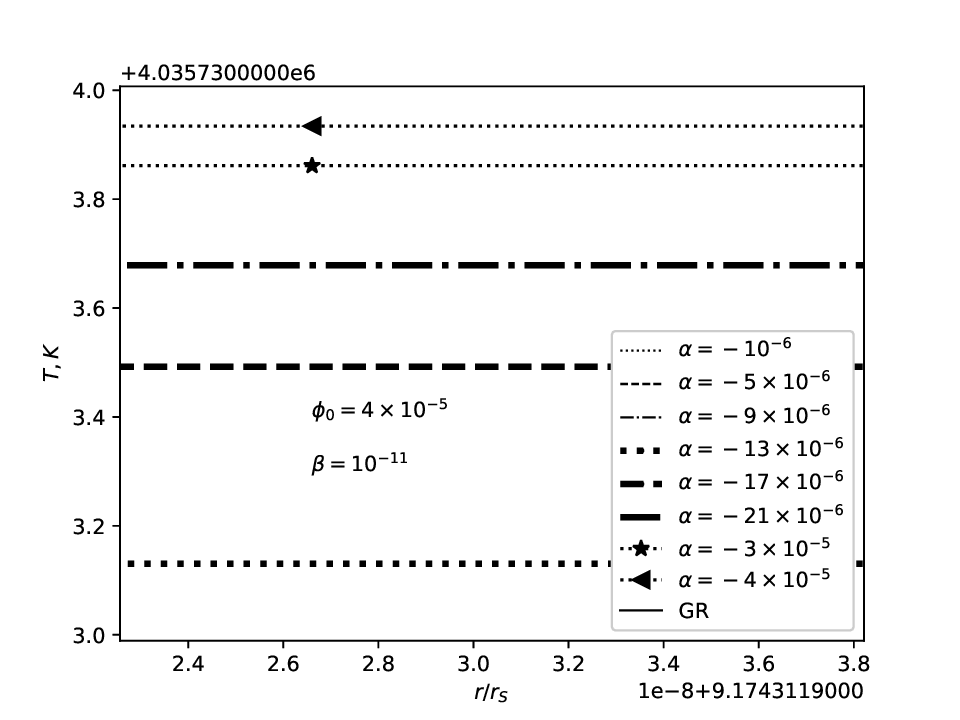}} \\ d)
			
		\end{minipage}
	\hfill
\begin{center}
\begin{minipage}[h]{.55\textwidth}
			{\includegraphics[width=\columnwidth]{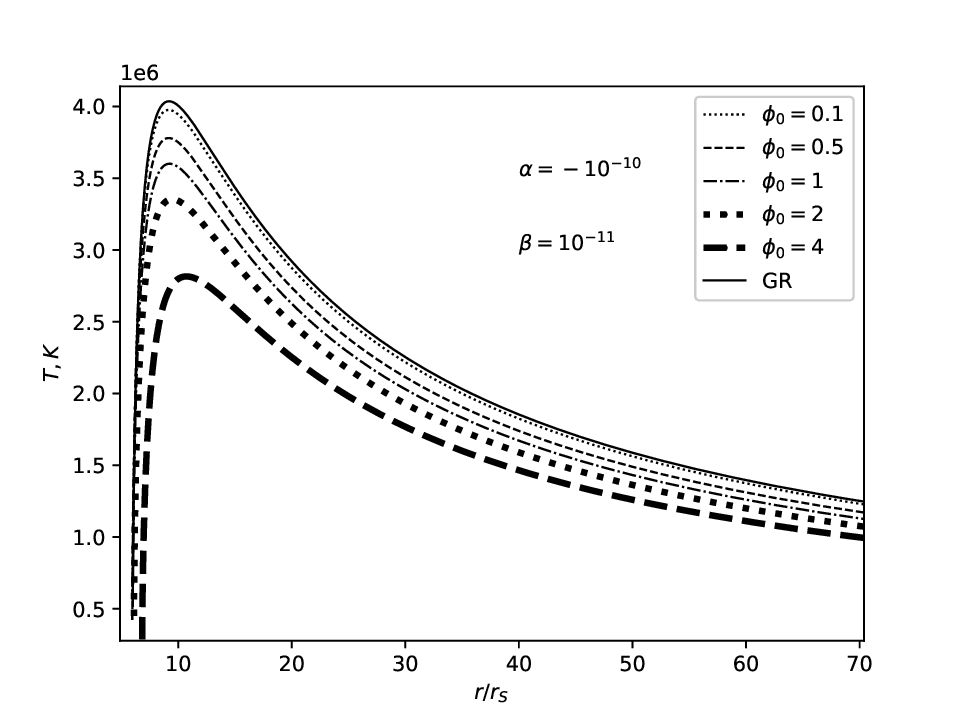}} \\ e)

	\end{minipage}
	\end{center}
		\caption{Hyggs-type potential case. The temperature distribution $T(r)$ of a disk around a static black hole with $\dot M = 2.21\times10^{18} g/s$ and $M=8.48 M_\odot$  as function of the normalized radial coordinate $r/r_s$. The connection  $m_\varphi^2=-\mu^2$ is taken into accaunt.  b), c), d) are zoom versions of a).}
		\label{fig:TV4}
	\end{figure*}

Temperature is related to energy flux through Stefan-Boltzmann equation. As in the case without potential, all conclusions that are true for the energy flux are also true for the effective temperature. All results for the temperarure in the case with potential are represented in Fig. \ref{fig:TV1}, \ref{fig:TV3}, \ref{fig:TV4}.


\subsubsection{Efficiency.} 
\begin{figure*}
		\begin{minipage}[h]{0.3\textwidth}
			\includegraphics[width=\columnwidth]{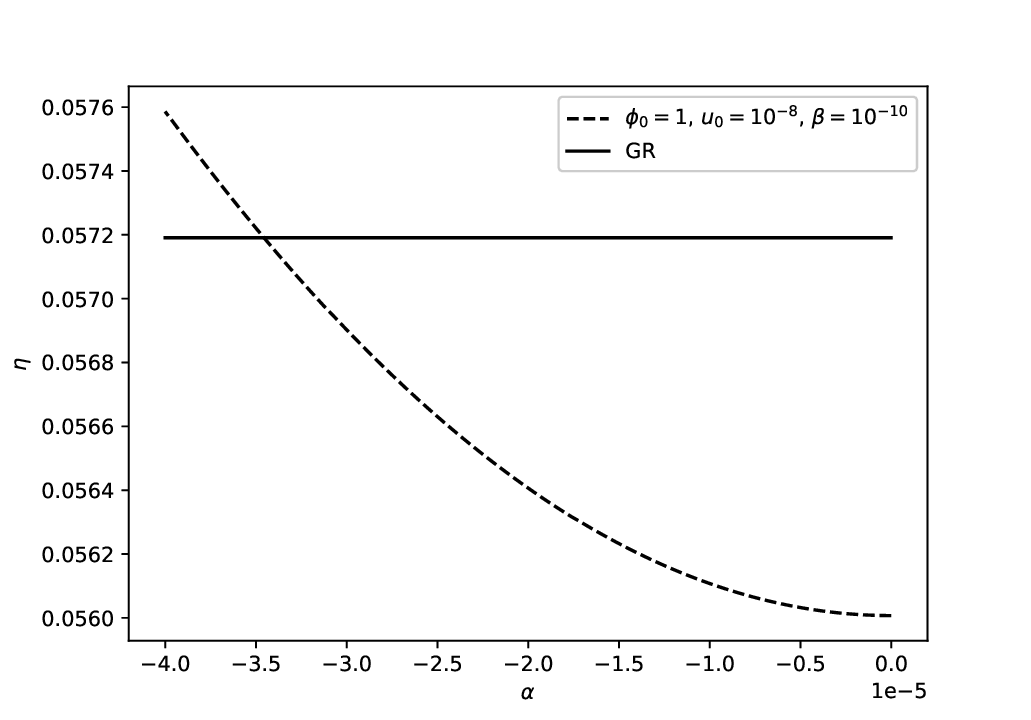} \\ a)
					\end{minipage}
\hfill
		\begin{minipage}[h]{.3\textwidth}
			{\includegraphics[width=\columnwidth]{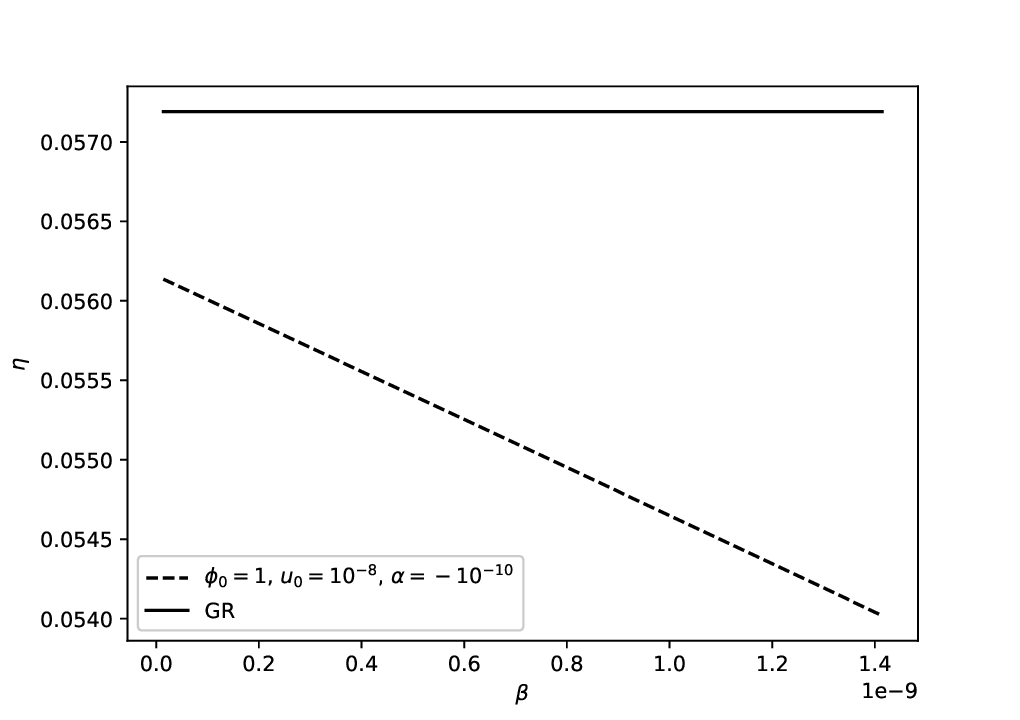}} \\ b)
			
		\end{minipage}
\hfill
		\begin{minipage}[h]{.3\textwidth}
			\includegraphics[width=\columnwidth]{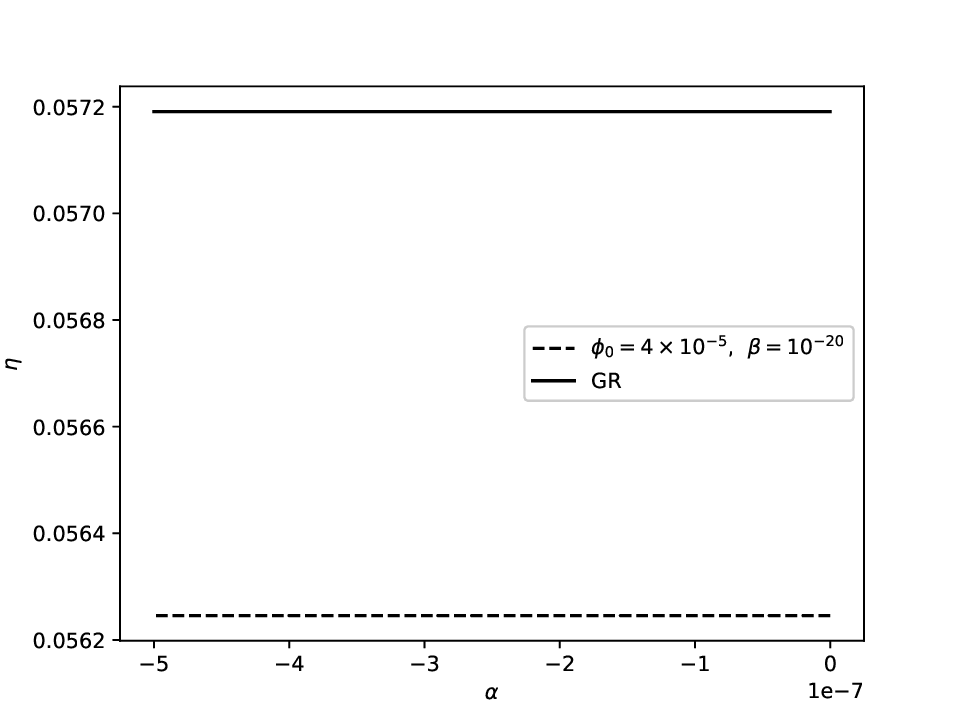} \\ c)
					\end{minipage}
\hfill

		\begin{minipage}[h]{.3\textwidth}
			{\includegraphics[width=\columnwidth]{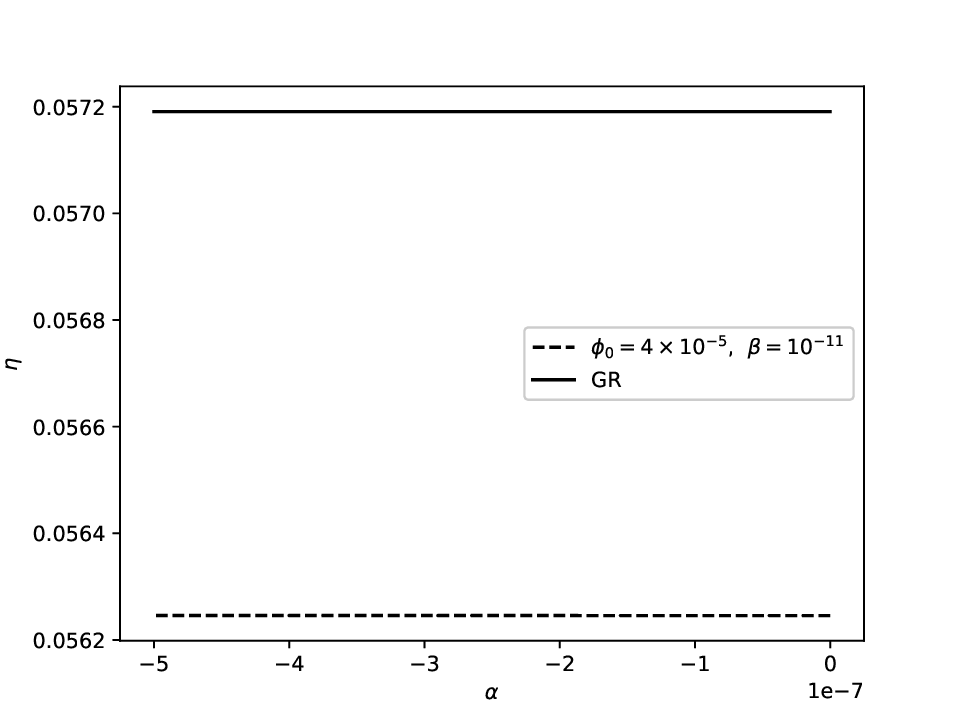}} \\ d)
			
		\end{minipage}
\hfill
		\begin{minipage}[h]{.3\textwidth}
			\includegraphics[width=\columnwidth]{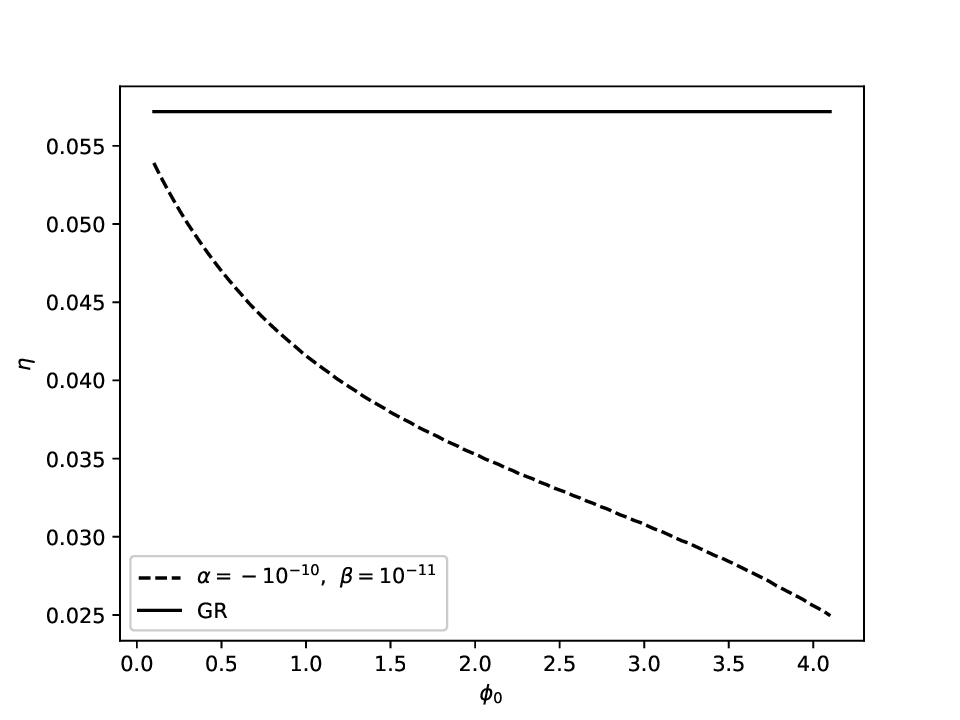} \\ e)
					\end{minipage}

		\caption{	Hyggs-type potential case. 	The efficiency for thin accretion disk around static black hole a)  for different values of $\alpha$  b)   for different values of $\beta$ ; c)   for different values of $\alpha$. The connections (\ref{hyggs}) and (\ref{u0V})  are taken into accaunt; d) 	for different values of $\alpha$. The connection $m_\varphi^2=-\mu^2$ is taken into accaunt; e) for different values of $\phi_0$. The connection  $m_\varphi^2=-\mu^2$ is taken into accaunt. }
		\label{fig:effV}
	
	\end{figure*}

For all previously considered cases, we also evaluated the effect of changing various parameters on the efficiency (see Fig. \ref{fig:effV}). 
\begin{enumerate}
\item As the modulus of $\alpha$ increases, the efficiency also increases. At values of $\alpha<-3.5\times10^{-5}$, the efficiency exceeds the Schwarzschild predictions. 
\item With an increase in  $\beta$  and a fixed $\alpha$, a decrease in efficiency is observed. Moreover, the Schfwarzschild value cannot be achieved for any value of  $\beta$  in this case. 
\item In this case, neither when changing $\alpha$  nor when changing $\beta$ are there any changes in the efficiency value. It remains  constant and is 98\% of the value predicted by Schwarzschild solution. 
\item In the case of $m_\phi^2 = -\mu^2$, we  obtain the same result as in the previous one. 
\item However, when considering large values of $\phi_0>1$ and taking into account $m_\phi^2 = -\mu^2$, we observe a significant decrease in efficiency, which can be less than 40\% of the Schwarzschild prediction.
\end{enumerate}
\section{Discussion}
In this paper,  the properties of thin accretion disks around static spherically symmetric black holes in hybrid metric-Palatini gravity are investigated. As a foundation for our research, we use the numerical black hole solution  obtained in the article \cite{Danila2019}. To study accretion properties, the steady-state Novikov-Thorne model is employed and the observational data of the system MAXI J1820+070 are used.  In this paper, we consider two types of solutions: first, we study a solution without a potential $V=0$, and then we take a Higgs-type potential $V=-\frac{\mu^2}{2}\phi^2+\frac{\zeta}{4}\phi^4$. As characteristics of the accretion disk, we numerically obtain the energy flux, temperature distribution, the emission spectra and the efficiency. The numerical black hole solution, which is derived in the article \cite{Danila2019}, has a certain set of free parameters. This set is determined, among other things, by the presence of the potential. In the case without a potential $V=0$, these parameters include the initial value of the scalar field $\phi_0$ and its derivative $u_0$. In the case with a Hyggs-type potential there are two additional parameters: $\alpha$ and $\beta$.

In the case $V=0$, we find the following features of the accretion disks in the HMPG. Results, which are close to Schwarzschild ones, can be obtained when we take sufficiently large values of $\phi_0$ for large $u_0$ or small values of $u_0$ for small $\phi_0$. However, the former result seems unrealistic, since the scalar field should take its background value at a large distance from the black hole. This value is significantly less than unity \cite{Leanizbarrutia2017},  \cite{Dyadina2019}, \cite{Dyadina2018}. Therefore, $\phi_0>1$  looks unnatural. 

Another approach to selecting the initial parameters $\phi_0$ and $u_0$ arises from post-Newtonian analysis. Far from the black hole, where we take the values of the free parameters, the gravitational field is weak, allowing us to consider the post-Newtonian expansion. Within the post-Newtonian analysis, the scalar field is considered as the sum of the background value and its perturbation $\phi=\phi_0+\varphi$. The background value $\phi_0$ is a constant quantity, unlike the perturbation $\varphi$. Thus, taking the derivative of the scalar field $\phi$ with respect to the distance at infinity, we simply obtain the value of $u_0$. As a result, we derive the connection eq. (\ref{u0}) between $u_0$ and $\phi_0$. If such a connection is established and the values of $\phi_0$ are taken within the limits set by the Cassini experiment \cite{Cassini2003}, then we obtain the curves for the energy flux, temperature distribution and the emission spectra that practically do not deviate from the Schwarzschild results (see Figures (\ref{fig:flux1}d), (\ref{fig:lum1}d)). In this case, $u_0$ takes a small values ($\sim10^{-11}$), which once again speaks in favor of choosing small values of the initial parameters due to their naturalness.

In the case of a Higgs-type potential, the theory includes four free parameters: $u_0$, $\phi_0$, $\alpha$ and $\beta$. Parameters $\alpha$ and $\beta$ are woven into the structure of the potential itself. At first, we consider all parameters as independent quantities. A large modulus $\alpha$ and sufficiently large $\phi_0>1$ lead to a situation, where the maximum energy flux and the emission spectrum can exceeds the Schwarzschild prediction. This situation contradicts the following idea:  large values of $\phi_0$ are not consistent with data obtained from other observations \cite{Leanizbarrutia2017},  \cite{Dyadina2019}, \cite{Dyadina2018}.  This fact leads us to conclusion that this combination is unrealistic. Further, with an increase in the $\beta$ parameter, a decrease in the energy flux maximum is observed, although not significant. No $\beta$ parameter yields a curve that exceeds the Schwarzschild prediction.

Then, we once again consider a set of parameters based on the limitations obtained from experiments in the solar system. We also use the relationship between the parameters resulting from the post-Newtonian expansion. In this case, all parameters turn out to be small. With an increase in the modulus $\alpha$, a decrease in the maximum of energy flux is observed.

The last case  is based on an analogy with particle physics. We assume that the mass of the scalar field $m_\phi^2=-\mu^2$. The relationship (\ref{u0V}) between $\phi_0$, $u_0$ and the mass of the scalar field $m_\phi$ is preserved. In this case, it is found that even at large values of $\alpha$, the energy flux, temperature distribution and the emission spectra would still be less than Schwarzschild predictions. However, if the parameter $\phi_0$ is increased, then at sufficiently large values of $\phi_0$, the energy flux, temperature distribution and the emission spectra begin to decrease significantly, and the difference can already exceed 80\% at $\phi_0=4$.

Another important characteristic of the accretion process  is the efficiency.  This quantity demonstrates the ability of the central body to convert rest mass into outgoing radiation. In the case without potential, the efficiency values closest to the Schwarzschild prediction can be achieved when both $u_0$ and $\phi_0$ are taken to be either large or small. The most realistic result is obtained by taking into account the relationship (\ref{u0}) between $\phi_0$ and $u_0$, and $\phi_0$ is taken within the constraints imposed from the solar system. When we consider a model with a Hyggs-type potential, the results for efficiency are the same as for other accretion characteristics. The only case where the efficiency exceeds the Schwarzschild prediction is when $\alpha$, $\phi_0$ and $u_0$ are all large. An interesting situation is observed when we take  small values of $\phi_0$ and $u_0$. Regardless of the type of connection between these parameters, the efficiency becomes constant. At the same time, for any values of the parameters $\alpha$ and $\beta$, it does not reach the Schwarzschild prediction if $\phi_0$ is taken to be small. In the case of large values of $\phi_0>1$ and the presence of a relationship (\ref{u0V}) between the parameters, there is a significant deviation from Schwarzschild predictions.

In this paper, we investigate the case of stellar-mass black holes. This is done because we select system MAXI J1820+070 with the smallest Kerr parameter $(a=0.14)$, since we consider a static spherically symmetric solution in HMPG. However, it is worth noting that our conclusions are also true for supermassive black holes. The general character of all curves is preserved, however, in the case of supermassive black holes, the energy flux has different order of maximum values. This quantity is less, and the difference can reach about $10^{15}$ $erg/(s \cdot cm^2)$ and even more, depending on the masses of the black holes. However, the gap between the predictions of HMPG and GR remains the same, regardless of the data of the binary system. Similar changes are observed for temperature distribution. In the case of luminosity, we find that the maximum values are observed at lower frequencies in the case of supermassive black holes. We can also see that the maximum itself is several orders of magnitude smaller than in the case of stellar-mass black holes. In addition, the efficiency of accretion does not depend on the mass of the black hole.

\section{Conclusions}
In this work, we have studied thin accretion disks around black holes in hybrid metric-Palatini gravity. In this investigation, we relied on a numerical static spherically symmetric solution \cite{Danila2019}. As a result, we have obtained the energy flux, temperature distribution, emission spectrum and efficiency for such black holes. We have shown that, in HMPG, accretion disks around static spherically symmetric black holes are colder and less luminous than in GR. This distinguishes HMPG from metric f(R)-theory, where thin accretion disks around such black holes are hotter and more luminous \cite{Perez2013}.

In this paper, we have considered various combinations of the free parameters of the hybrid metric-Palatini gravity in order to determine its viability. One case shows the excess in the flux,  temperature  and luminosity relative to the values predicted by the Schwarzschild solution: when we consider a model with a Higgs-type potential and with large values of parameters $\alpha$, $\phi_0$ and $u_0$. However, such a set of parameters  is not consistent with the limitations imposed by other experiments on the background value of the scalar field (the value away from the black hole) \cite{Leanizbarrutia2017, Dyadina2018, Dyadina2019, Avdeev2020}.  Thus we assume that the set of large values of $\phi_0$, $u_0$ and $\alpha$ is unrealistic. All other sets of parameters demonstrate that the Schwarzschild values are not reached.  

The hybrid metric-Palatini gravity has another important consequence. Within the restrictions imposed on the parameters of the theory by other methods \cite{Leanizbarrutia2017, Dyadina2018, Dyadina2019, Avdeev2020}, this model shows its full viability, and the results for the energy flux, temperature, and emission spectrum are close to GR. Another advantage of HMPG is that realistic accretion regimes are implemented in a wide range of parameters, without their fine tuning. In addition, the existence of realistic accretion regimes indicates the adequacy of the numerical black hole solution  obtained in the article \cite{Danila2019}. In this work, the authors demonstrate the possibility of the existence of spherically symmetric black holes in a wide range of parameter values ($0.5<\phi_0<8$, $4\times10^{-9}<u_0<5.12\times10^{-7}$). However, this range is not consistent with limitations from other observations ($\phi_0<4\times10^{-5}$). We show that such a solution is also possible in the case of parameter values consistent with observations obtained from the solar system and binary  pulsars \cite{Leanizbarrutia2017, Dyadina2018, Dyadina2019, Avdeev2020}. Thus, additional confirmation of the viability of this solution is the presence of realistic accretion regimes with such a natural set of parameters.

Accretion disks serve as excellent platforms for testing theories of gravity, as well as for constraining them. One of the most popular methods of testing such theories is the continuum-fitting method \cite{Zhang_1997}. This method relies on the fact that the thermal spectrum of the disk depends on the background metric and on the movement of massive particles within the accretion disk, as well as on the path of photons from the emission point in the disk to the point of detection. Importantly, these dependencies do not involve atomic physics \cite{Bambi_2021}. That is, all differences are determined by the underlying metric. To investigate whether continuum observations can constrain models of gravity, it is necessary to theoretically estimate the luminosity and compare it with observations using the minimum chi-square $(\chi^2)$ approach \cite{mohaddese}. Additionally, another significant test of gravitational theories using accretion disk systems involves examining predictions related to iron line broadening \cite{LU_2003}. The fluorescent line of iron, at an energy of 6.4 keV, allows to investigate the space-time metric near accreting compact objects. The theory of gravity influences the expected shape of this line, making it an excellent test of the theory, as it allows the comparison between the predictions and observed data. This task is especially relevant given the latest data obtained from X-ray sources such as NuSTAR, RXTE, Suzaku, Swift, and XMM-Newton. Moreover, it is highly probable that the resulting theoretical predictions can be tested using next-generation detectors like ATHENA.

It is important to emphasize that all the conclusions drawn in this work apply to both stellar-mass and supermassive black holes. The only differences observed are in the values of the maximum energy flux, luminosity, and temperature, which are lower for supermassive black holes. Additionally, the emission spectrum curves shift to the left with increasing mass, indicating that the maximum luminosity occurs at lower frequencies. Furthermore, the mass of the black hole does not in any way affect the efficiency of accretion.

In this paper, we consider the case of a static spherically symmetric black hole. This is a crucial first step in researching accretion in hybrid metric-Palatini gravity. Despite the fact that the probability that this type of black holes is realized in nature is extremely small, it's vital to understand the potential existence of adequate accretion regimes for this type. The next step of this research will be to consider accretion in case of Kerr-type black holes.  It will allow to compare predictions of HMPG with observations. This study will shed light on the realism of the theory and serve as the basis for imposing restrictions on free parameters of the hybrid metric-Palatini gravity, including through the
study of accretion disks around rotating black holes. This is a subject for future research.

\begin{acknowledgements}
The authors thank G. V. Lipunova for discussions and comments on the topics of this paper. P.I. Dyadina acknowledges support from Russian Science Foundation grant 22-72-00022.
\end{acknowledgements}

\textbf{Data Availability Statement} This manuscript has associated data
in a data repository. [Authors’ comment: All observational data
used in the article are presented in the work https://doi.org/10.3847/
9421538-4357/ac07a9.]

\section*{Declarations}

\textbf{Conflict of interest} The authors declare that they have no known competing financial interests or personal relationships that could have appeared to influence the work reported in this paper.

\bibliographystyle{spphys}
\bibliography{accretion_epjc.bib}

\end{document}